\pdfoutput=1
\documentclass[%
  letterpaper,%
  twocolumn,%
  amsmath,%
  amssymb,%
  amsfonts,%
  superscriptaddress,%
  groupedaddress,%
  prb,%
  showpacs,%
]{revtex4}

\usepackage{bm,graphicx,hyperref}
\hypersetup{%
  breaklinks = {true},
  citecolor = {blue},
  colorlinks = {true},
  linkcolor = {red},
}

\newcommand{\mbf}{\mathbf}
\newcommand{\ie}{\emph{i.e.}}
\newcommand{\Eqref}[1]{Eq.~\eqref{#1}}
\newcommand{\Fref}[1]{Fig.~\ref{#1}}

\begin{document}

\title{Pseudomagnetic fields in graphene nanobubbles of constrained geometry: A 
molecular dynamics study}

\author{Zenan~Qi%
\footnote{Electronic address: zenanqi@bu.edu}%
}
\affiliation{Department of Mechanical Engineering, Boston University, Boston, MA 
02215}

\author{Alexander~L.~Kitt%
\footnote{Electronic address: alkitt@bu.edu}%
}
\affiliation{Department of Physics, Boston University, 590 Commonwealth Ave, 
Boston, Massachusetts 02215, USA}

\author{Harold~S.~Park%
\footnote{Electronic address: parkhs@bu.edu}%
}
\email[Corresponding author: ]{parkhs@bu.edu}
\affiliation{Department of Mechanical Engineering, Boston University, Boston, MA 
02215}

\author{Vitor~M.~Pereira%
\footnote{Electronic address: vpereira@nus.edu.sg}%
}
\email[Corresponding author: ]{vpereira@nus.edu.sg}
\affiliation{Graphene Research Centre \& Department of Physics, National 
University of Singapore, 2 Science Drive 3, Singapore 117542}

\author{David~K.~Campbell%
\footnote{Electronic address: dkcampbe@bu.edu}%
}
\affiliation{Department of Physics, Boston University, 590 Commonwealth Ave, 
Boston, Massachusetts 02215, USA}

\author{A.~H.~Castro~Neto%
\footnote{Electronic address: neto@bu.edu}%
}
\affiliation{Department of Physics, Boston University, 590 Commonwealth Ave, 
Boston, Massachusetts 02215, USA}
\affiliation{Graphene Research Centre \& Department of Physics, National 
University of Singapore, 2 Science Drive 3, Singapore 117542}
\affiliation{Department of Electrical and Computer Engineering, National 
University of Singapore, 4 Engineering Drive 3, Singapore 117583}

\date{\today}

\pacs{81.05.ue, 73.22.Pr, 71.15.Pd, 61.48.Gh}


\begin{abstract}

Analysis of the strain-induced pseudomagnetic fields generated in graphene 
nanobulges under three different substrate scenarios shows that, in addition to 
the shape, the graphene-substrate interaction can crucially determine the 
overall distribution and magnitude of strain and those fields, in and outside 
the bulge region. We utilize a combination of classical molecular dynamics, 
continuum mechanics, and tight-binding electronic structure calculations as an 
unbiased means of studying pressure-induced deformations and the resulting 
pseudomagnetic field distribution in graphene nanobubbles of various geometries. 
 The geometry is defined by inflating graphene against a rigid aperture of a 
specified shape in the substrate. The interplay among substrate aperture 
geometry, lattice orientation, internal gas pressure, and substrate type is 
analyzed in view of the prospect of using strain-engineered graphene 
nanostructures capable of confining and/or guiding electrons at low energies. 
Except in highly anisotropic geometries, the magnitude of the pseudomagnetic 
field is generally significant only near the boundaries of the aperture and 
rapidly decays towards the center of the bubble because under gas pressure at 
the scales considered here there is considerable bending at the edges and the 
central region of the nanobubble displays nearly isotropic strain. When 
the deflection conditions lead to sharp bends at the edges of the bubble, 
curvature and the tilting of the $p_z$ orbitals cannot be ignored and 
contributes substantially to the total field. The strong and localized nature of 
the pseudomagnetic field at the boundaries and its polarity-changing profile can 
be exploited as a means of trapping electrons inside the bubble region or of 
guiding them in channel-like geometries defined by nano-blister edges. However, 
we establish that slippage of graphene against the substrate is an important 
factor in determining the degree of concentration of pseudomagnetic fields in or around the bulge 
since it can lead to considerable softening of the strain gradients there. The 
nature of the substrate emerges thus as a decisive factor determining the 
effectiveness of nanoscale pseudomagnetic field tailoring in graphene.
\end{abstract}

\maketitle

\section{Introduction}
Since the discovery of a facile method for its 
isolation, graphene ~\cite{novoselovNATURE2005},  the simplest two-dimensional
crystal, has attracted intense
attention not only for its unusual physical
properties~\cite{geimNM2007,netoRMP2009,hanPRL2007,seolSCIENCE2010},
but also for its potential as the basic building block for
a wealth of device applications.
There exist key limitations that appear to restrict the
application of graphene for all-carbon electronic circuits: one such
limitation is that graphene, in its pristine form, is well known to be
a semi-metal with no band gap~\cite{netoRMP2009}.  A highly active
field of study has recently emerged based on the idea of applying
mechanical strain to modify the intrinsic response of electrons to external 
fields in graphene ~\cite{GuineaNaturePhy2010,qiNL2013,TomoriAPE2011}. 
This includes the strain-induced generation of spectral 
(band) gaps and transport gaps, which suppress conduction at small densities.
In this context, several groups ~\cite{GuineaPRB2010,PereiraPRL2009,GuineaPTRSMPES2010,GuineaPRB2008,
GuineaPRB2011,GuineaNaturePhy2010,KimPRB2011,YehSS2011,YangJPCM2011,
KittPRB2012,KittErratum:2013,YueJAP2012}
have employed continuum mechanics 
coupled with effective
models of the electronic dynamics to study the generation of
pseudomagnetic fields (PMFs) in different graphene geometries and
subject to different deformations.  The potential impact of strain
engineering beyond the generation of bandgaps has also attracted
tremendous 
interest~\cite{PereiraPRL2009,PereiraPRL2010,AbaninPRL2012,WangPRB2011}.

\citet{PereiraPRB2009} showed that a band gap will not emerge under
simple uniaxial strain unless the strain is larger than roughly 20\,\%.
This theoretical
prediction, based on an effective
tight-binding model for the electronic structure, has been
subsequently confirmed by various more elaborate \emph{ab-initio}
calculations \cite{abinitio-Ping,abinitio-Farjam,abinitio-Son}. 
The robustness of the gapless state arises because
simple deformations of the lattice lead only to local changes of the
{\it position} of the Dirac point with respect to the undeformed lattice
configuration \cite{Kane:1997,Suzuura}
and to  anisotropies in the Fermi surface and Fermi velocity
\cite{Pereira-OpticalStrain:2010}. The shift in the position of the
Dirac point is captured, in the low-energy, two-valley, Dirac
approximation, by a so-called pseudomagnetic vector potential 
and resulting pseudomagnetic field (PMF) that
arises from the strain-induced perturbation of the tight-binding
hoppings \cite{Suzuura}. As a result, electrons react to
mechanical deformations in a way that is analogous to their behavior
under a real external magnetic field, except that overall time-reversal
symmetry is preserved, since the PMF has opposite signs in the two
time-reversal related valleys \cite{netoRMP2009}.

\citet{GuineaNaturePhy2010} found that nearly homogeneous PMFs could
be generated in graphene through triaxial stretching, but the
resulting fields were found to be moderate, unless relatively large
(\ie,  $>$10\,\%) tensile strains could be applied.  Unfortunately, such
large planar tensile strains have not been experimentally realized in
graphene to date. This is arguably attributed to the record-high
tensile modulus of graphene and the unavoidable difficulty in
effectively transferring the required stresses from substrates to this
monolayer crystal \cite{Gong}.

It is thus remarkable that recent experiments report the detection of
non-uniform strain distributions in bubble-like corrugations that
generate PMFs locally homogeneous enough to allow the observation of
Landau quantization by local tunneling spectroscopy. The magnitude of
the PMFs reported from the measured Landau level spectrum reaches
hundreds (300 to 600)
of Teslas~\cite{LevyScience2010,LuNatureComm2012}, providing a
striking glimpse of the impact that local strain can potentially have
on the electronic properties.  A difficulty with these experiments is
that, up to now, such structures have been seen and/or generated only
in contact with the metallic substrates that are used in the synthesis
of the sample.  This is an obstacle, for example, to transport
measurements, since this would require the transfer of the graphene sheet
to another substrate, thereby destroying the favorable local strain
distribution. In addition, a systematic study of different graphene
bubble geometries and substrate types, which could reveal the subtleties that 
different geometries bring to the related strain-induced PMFs
has not been reported.  Furthermore, most previous studies of the
interplay between strain and electronic structure in graphene
have addressed the deformation problem from an analytic continuum mechanics 
point of view, with the exception of a few recent computational
studies~\cite{Neek-AmalPRB2012b,Neek-AmalPRB2012c}.

It is in this context that we report here results from classical
molecular dynamics (MD) simulations of strained graphene nanobubbles
induced by gas pressure. The MD simulations are used to complement
and compare continuum mechanics approaches to calculating
strain, in order to examine the pressure-induced PMFs in ultra-small
graphene nanobubbles of diameters on the order of 5\,nm.
Controlled synthesis of such small strained nanobubbles has gained
impetus following the recent experiments by Lu {\it et al.}
\cite{LuNatureComm2012}.
Our aim is to use an unbiased calculation for the mechanical response
of graphene at the atomistic level, on the basis of which we can
(i) extract the relaxed lattice configurations without any
assumptions;
(ii) calculate the PMF distribution associated with different nanobubble
geometries; 
(iii) discuss the influence of substrate and aperture shape on PMF distribution;
(iv) identify conditions under which explicit consideration of the
curvature is needed for a proper account of the PMFs.

We first describe the simulation methodology that was employed to determine the 
atomic displacements from which the strain tensor, modified electronic 
hopping amplitudes, and PMFs can be obtained.  This is followed by numerical 
results of the strain-induced PMFs for different graphene nanobubble 
geometries in a simply clamped scenario. We next discuss the considerable 
importance of the substrate interaction and, finally, analyze the relative 
contributions of orbital bending and bond stretching to the total PMF.

%
\begin{figure} \centering
\includegraphics[scale=0.4]{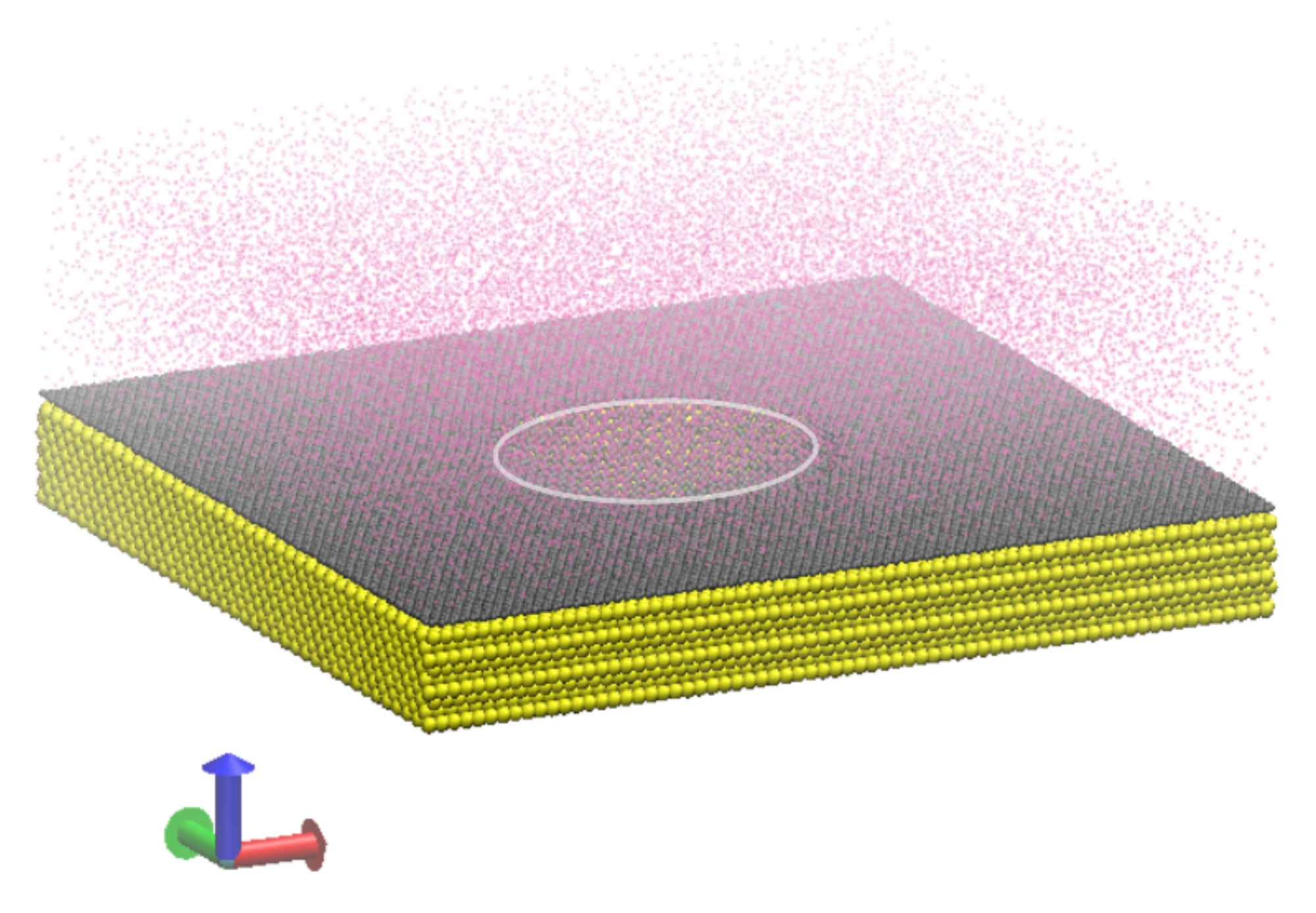}
\caption{(Color online) Illustration of the strategy employed in our studies to 
generate nanobubbles by pressurizing graphene through a predefined substrate 
aperture. The picture shows one of the actual simulation cells used in our MD 
computations.
In gold, gray and red  colors are represented, respectively, the Au substrate, 
the graphene sheet and the Ar atoms. A hole is carved in the Au substrate 
(perimeter outlined), and its perimeter geometry determines the shape of the 
resulting graphene bubble. Visualization is performed using VMD~\cite{Humphrey1996}.} 
\label{fig:schematic}
\end{figure}

\section{Simulation methodology}

Recent experiments have shown that graphene nanobubbles smaller than
10\,nm can be prepared on metallic substrates, and that large
PMFs in the hundreds of Tesla result from the locally induced
non-homogeneous strain ~\cite{LevyScience2010,LuNatureComm2012}.
Because such small nanobubbles can be directly studied using
classical MD simulations, we employ MD to obtain the deformed graphene
bubble configurations due to an externally applied pressure.The
atomistic potentials that describe the carbon-carbon interactions have
been extensively investigated and, hence, graphene's nano-mechanics
can be simulated without any particular bias, and to a large accuracy
within MD.  Once the deformation field is known from the simulations,
we obtain the strain distribution in the inflated nanobubble,
finally followed by a continuum gauge field approach to extract the
resulting PMF distribution
~\cite{netoRMP2009,GuineaNaturePhy2010,GuineaPRB2010,
GuineaPTRSMPES2010,GuineaPRB2008,GuineaPRB2011}. 

\subsection{Details of the MD simulations}

Our MD simulations were done with the Sandia-developed open source code
LAMMPS~\cite{plimptonLAMMPS,PlimptonJCP1995}. The graphene nanobubble system 
consisted of three parts, as illustrated in Fig.~\ref{fig:schematic}:  a 
graphene monolayer, the hexagonal (111) surface of an FCC gold substrate, and 
argon gas which was used to inflate the graphene bubble.  
We used the AIREBO potential~\cite{stuartJCP2000} 
to describe the C-C interactions, as this potential has been shown 
to describe accurately the various carbon interactions including 
bond breaking and reforming~\cite{zhaoJAP2010,qiNANO2010}. The substrate-graphene
and gas-graphene interactions were modeled by a standard 12-6 Lennard-Jones 
potential: 
\begin{equation}
  V(r_{ij}) =
  4\,\epsilon_{ij}\left[\left(\frac{\sigma_{ij}}{r_{ij}}\right)^{12}
  -\left(\frac{\sigma_{ij}}{r_{ij}}\right)^{6}\right],
  \label{eq:lj}
\end{equation}
where $r_{ij}$ represents the distance between the $i$-th carbon and the $j$-th 
gold atom.

The dimension of the simulation box was 20$\times$20$\times$8\,nm$^3$,
and the substrate was comprised of Au atoms with a thickness of 2\,nm, or 
about 2.5 times the cutoff distance of the interatomic 
potential~\cite{Neek-AmalPRB2012a}.
Apertures of different shapes (viz. triangle, rectangle, square, pentagon,
hexagon, and circle) were ``etched'' in the center of the substrate to
allow the graphene membrane to bulge inwards due to the pressure
exerted by the Ar gas.  The whole system
was first relaxed for 50\,ps, at which time the Ar gas was pushed downward
(as in a piston) to exert pressure on the graphene monolayer, causing
it to bulge inward in the shape cut-out from the gold substrate. The system is 
then allowed to equilibrate again under the increased gas pressure. All
simulations were carried out at room temperature (300\,K) using the
Nose-Hoover thermostat~\cite{hooverPRA1985}. The choice 
of Ar in our calculations is not mandatory. Substitution with other molecular 
species should pose no difficulty, the same being true regarding the
substrate, as shown previously in references 
\onlinecite{Neek-AmalPRB2012a} and \onlinecite{Neek-AmalPRB2012c}. 

To elucidate the effect of different substrates on the PMF distributions in the 
nanobubbles, we perform MD simulations with two different substrates, in 
addition to performing the simulations with fixed edges and no substrate.  
Specifically, we used both Au and Cu\,(111) substrates, where the detailed 
parameters and descriptions will be discussed in later sections.

After obtaining the graphene bubble, we held the pressure constant for 10\,ps to 
achieve thermal equilibrium. We note that during the 
entire simulation no gas molecules leaked away from the system, 
which again demonstrates the experimentally observed atomic impermeability of 
monolayer graphene~\cite{NairScience2012,BunchNanoLett2008}.  

Our simulations are close in spirit to the experiments reported in reference 
\onlinecite{Bunch:2011}, but targeting smaller hole apertures due to computation 
limitation. We note that this method of using gas-pressure to generate the
graphene nanobubbles is different from the situations explored in the
recent experiments that focus on the PMF
distribution~\cite{LevyScience2010,LuNatureComm2012}. However, it is
in some ways more controllable due to the utilization of a substrate
with a distinct pattern coupled with externally applied pressure to
force graphene through the patterned substrate to form a bubble with
controllable shape and height.  

The final (inflated bubble) configuration gives us the basic ingredients needed 
to extract the strain distribution in the system, as well as the perturbed 
electronic hopping amplitudes. To calculate the strain directly from the 
displaced atomic positions we employ what we shall designate as the
\emph{displacement approach}. We note that a previous study 
\cite{KlimovScience2012} used a \emph{stress}
approach for a similar calculation. However, the stress approach fails to 
predict reasonable results in our case, which we attribute to the
inability of the virial stresses to properly convey the total stress
at each atom of the graphene sheet when the load results from interaction with 
gas molecules.  Furthermore, in the stress approach one assumes a planar (and, 
in addition, usually linear) stress-strain constitutive relation which leads 
to errors when large out-of-plane deformations arise, as in the case of the 
nanobubbles.  Further details on the strain calculation are given in appendix 
\ref{ap:disp-vs-stress}.

\subsection{Displacement approach to calculate strain}%
\label{sec:displacement}

In continuum mechanics the infinitesimal strain tensor is written in
Cartesian material coordinates ($X_i$) as
\begin{equation} \label{eq:straindisp}
\epsilon_{ij}=\frac{1}{2}\left(\frac{\partial {u_i}}{\partial {X_j}} +
\frac{\partial {u_j}}{\partial 
{X_i}}\right)+\frac{1}{2}\left(\frac{\partial{u_k}}{\partial{X_i}}\frac{\partial
{u_k}}{\partial{X_j}}\right)
\!\!.
\end{equation}
To utilize Eq.~\ref{eq:straindisp}, it is clear that the displacement
field must be obtained such that its derivative can be evaluated to
obtain the strain.  
In order to form a linear interpolation scheme using finite 
elements~\cite{hughes1987}, we exploit the geometry of the lattice and
mesh the results of our MD simulation of the deformed graphene bubble using 
tetrahedral finite elements defined by the positions of four atoms:  the atom
of interest (with undeformed coordinates $\mbf{R}_0$), and its three
neighbors (with undeformed coordinates $\mbf{R}_1$, $\mbf{R}_2$,
$\mbf{R}_3$).  After deformation, the new positions of
the atoms are $\mbf{r}_0$, $\mbf{r}_1$, $\mbf{r}_2$ and $\mbf{r}_3$,
respectively.  To remove spurious rigid body translation and rotation
modes, we took the atom of interest ($\mbf{R}_0$) as the reference
position, \ie,  $\mbf{r}_0$ = $\mbf{R}_0$.  The displacement of its
three neighbors could then be calculated, and subsequently the
components of the strain tensor $\epsilon_{ij}$ were obtained by
numerically evaluating the derivative of the displacement inside the
element.

\subsection{Pseudomagnetic fields}

Non-zero PMFs arise from the non-uniform displacement in the 
inflated state. These PMF reflect the physical perturbation that the
electrons near the Fermi energy in graphene feel as a result of the
local changes in bond length. It emerges straightforwardly in the
following manner. Nearly all low-energy electronic properties and
phenomenology of graphene are captured by a simple single orbital
nearest-neighbor tight-binding (TB) description of the $\pi$ bands in
graphene \cite{netoRMP2009}. In second quantized form this
tight-binding Hamiltonian reads
\begin{equation}
  H = - \sum_{i,\mbf{n}} t \bigl(\mbf{r}_i,\mbf{r}_i+\mbf{n}\bigr)
  \, a^\dagger_{\mbf{r}_i} b^{\phantom{\dagger}}_{\mbf{r}_i+\mbf{n}}
  + \text{H. c.}
  ,
\end{equation}
where $t\bigl(\mbf{r}_i,\mbf{r}_i+\mbf{n}\bigr)$ represents the
hopping integral between two neighboring $\pi$ orbitals, $\mbf{n}$
runs over the three nearest unit cells, and
$a_{\mbf{r}_i}$($b_{\mbf{r}_i}$) are the destruction operators at the
unit cell $\mbf{r}_i$ and sublattice $A(B)$. In the undeformed lattice
the hopping integral is a constant:
$t\bigl(\mbf{r}_i,\mbf{r}_i+\mbf{n}\bigr)=t\bigl(\mbf{R}_i,\mbf{R}
_i+\mbf{n}\bigr)=t=2.7$\,eV. The deformations of the graphene lattice caused by 
the gas pressure impact the hopping amplitudes in two main ways. One  
arises from the local stretch that generically tends to move atoms farther 
apart from each other and, consequently, directly affects the magnitude of the 
hopping $t_{ij}$ between neighboring atoms $i$ and $j$, which is exponentially 
sensitive to the interatomic distance. The other effect is caused by the 
curvature induced by the out-of-plane deflection, which means that the hopping 
amplitude is no longer a purely $V_{pp\pi}$ overlap (in Slater-Koster notation) 
but a mixture of $V_{pp\pi}$ and $V_{pp\sigma}$. More precisely, one can 
straightforwardly show that the hopping between two $p_z$ orbitals 
oriented along the unit vectors $\mbf{n}_i$ and $\mbf{n}_j$ and a distance 
$\mbf{d}$ apart is given by \cite{Isacsson:2008,PereiraPRL2010}
\begin{multline}
 - t_{ij} = V_{pp\pi}(d) \, \mbf{n}_i\cdot\mbf{n}_j  \\
          + \frac{V_{pp\sigma}(d) - V_{pp\pi}(d)}{d^2} \,
            (\mbf{n}_i\cdot\mbf{d})(\mbf{n}_j\cdot\mbf{d})
  \label{eq:Hopping}
  \,.
\end{multline}
To capture the exponential sensitivity of the overlap integrals to the 
interatomic distance $d$ we model them by
\begin{subequations} \label{eq:VvsDistance}
\begin{align}
  V_{pp\pi}(d)    &= - t\,e^{-\beta (d/a-1)}, \\
  V_{pp\sigma}(d) &= + 1.7\, t\, e^{-\beta (d/a-1)},
\end{align}
\end{subequations}
with $a\simeq 1.42$\,\AA\ the equilibrium bond length in graphene. For static 
deformations a value $\beta \approx 3$ is seen to capture the distance 
dependence of $V_{pp\pi}(d)$ in agreement with first-principles calculations 
\cite{PereiraPRB2009,Pereira-OpticalStrain:2010}; we use the same decay constant 
$\beta$ for both overlaps, which is justified from a M\"ulliken perspective 
since the principal quantum numbers of the orbitals involved is the same 
\cite{Hanson:2003}.

In the undeformed state \Eqref{eq:Hopping} reduces to 
$-t_{ij}=V_{pp\pi}(a)\equiv -t$ and is, of course, constant in the entire 
system. But local lattice deformations cause 
$t\bigl(\mbf{r}_i,\mbf{r}_i+\mbf{n}\bigr)$ to fluctuate, which we can describe 
by suggestively writing $t\bigl(\mbf{r}_i,\mbf{r}_i+\mbf{n}\bigr) = t + \delta 
t\bigl(\mbf{r}_i,\mbf{r}_i+\mbf{n}\bigr)$. In the low energy (Dirac) 
approximation, the effective Hamiltonian around the point $\mbf{\pm K}$ in the 
Brillouin zone can then be written as \cite{Kane:1997,Suzuura}
\begin{equation}
  H_\text{eff}^{\pm\mbf{K}} = v_F\,
\bm{\sigma}\cdot\bigl(\mbf{p} \mp q\mbf{A}\bigr),
  \label{eq:HDirac}
\end{equation}
where $\hbar v_F = 3 t a/2$, $q$ represents the charge of the current carriers 
($q>0$ for holes and $q<0$ for electrons), and the Cartesian components of the 
pseudomagnetic 
vector potential $\mbf{A}=A_{x}\bm{e}_x+A_{y}\bm{e}_y$ 
are given explicitly in terms of the hopping perturbation by
\begin{equation}
  A_{x}(\mbf{R}) - i A_{y}(\mbf{R})
  = \frac{1}{q v_F}
  \sum_{\bm{n}} \delta t\bigl(\mbf{r},\mbf{r}+\mbf{n}\bigr)
  e^{i\bm{K}\cdot\bm{n}}
  \label{eq:Adef-full}
  .
\end{equation}
For nearly planar deformations (small out-of-plane vs in-plane displacement 
ratios and thus neglecting bending effects) $\delta t$ can be expanded in terms 
of the local displacement field and, consequently, can be cast in terms of the 
strain components. Orienting the lattice so that the zig-zag direction is 
parallel to $\bm{e}_x$ leads to
\begin{equation}
  A_{x}(\mbf{R}) - i A_{y}(\mbf{R})
  \simeq \frac{\hbar\beta}{2 q a}
  \bigl(\epsilon_{xx} - \epsilon_{yy} + 2i\,\epsilon_{xy}\bigr)
  ,
  \label{eq:Adef-strain}
\end{equation}
Since we are ultimately interested in the PMF, only the contributions to 
$\mbf{A}(\mbf{R})$ arising from the hopping modification are considered here, 
as they are the ones that survive after the curl operation 
\cite{deJuan:2013,KittPRB2012,KittErratum:2013,Barraza-Lopez:2013,Naumis:2013}; 
we also don't consider contributions beyond second order smallness ($\sim k\,
\epsilon$, $\sim k^2$, etc.).
In the planar strain situation the whole information about the electronic 
structure is reduced to the parameter $\beta = - \partial\log t(r) / 
\partial\log r\bigr|_{r=a}$.

From the coupling in \Eqref{eq:HDirac} where the effects of strain are 
captured by replacing $\mbf{p}\to\mbf{p}-q\mbf{A}$ it is clear that the local 
strain is felt by the electrons in the $\bm{K}$ valley in the same way as an 
external magnetic field would be. In particular, we can quantify this effect in 
terms of the PMF, which is defined as
\begin{equation}
  B = \partial_x A_{y}(\mbf{R}) - \partial_y A_{x}(\mbf{R}) 
  .
  \label{eq:B-def}
\end{equation}
This is the central quantity of interest in this work; in the subsequent 
sections the combined effects of gas pressure, hole geometry, and substrate 
interaction will be analyzed from the point of view of the resulting magnitude 
and space distribution of the PMF, $B$, obtained in this way. For definiteness 
we set $q=e$, $e$ being the elementary charge, which means that we shall be 
analyzing the PMF from the perspective of holes ($q>0$).
From an operational perspective, $B$ can be 
calculated directly from \Eqref{eq:Adef-full} by computing the hopping between 
all pairs of neighboring atoms in the deformed state, or from 
\Eqref{eq:Adef-strain} by calculating the strain components throughout the 
entire system as described in the previous section. The former strategy is here 
referred to as the \emph{TB approach}, and the latter as the \emph{displacement 
approach}, as per the definitions in section \ref{sec:displacement}. Our PMF 
calculations in the following sections are done by following the TB approach, 
except when we want to explicitly compare the results obtained with the 
two approaches. In those cases, such as in the next section or in Appendix 
\ref{ap:pmf-disp}, that will be explicitly stated. 

\section{Clamped graphene nanobubbles}\label{sec:clamped}

We first simulated an idealized system consisting only of Ar gas molecules and 
graphene, neglecting the interaction with the underlying substrate, and where 
we strictly fixed all carbon atoms outside the aperture region during 
simulation.  This provides a good starting point to understand how the shape of 
the substrate aperture affects the PMF distribution.  A similar system 
has been used in previous work~\cite{WangJAM2013}, as this corresponds to a 
continuum model with clamped edges~\cite{GuineaPRB2008,KimPRB2011}.

\begin{figure*} \centering
\includegraphics[width=0.85\textwidth]{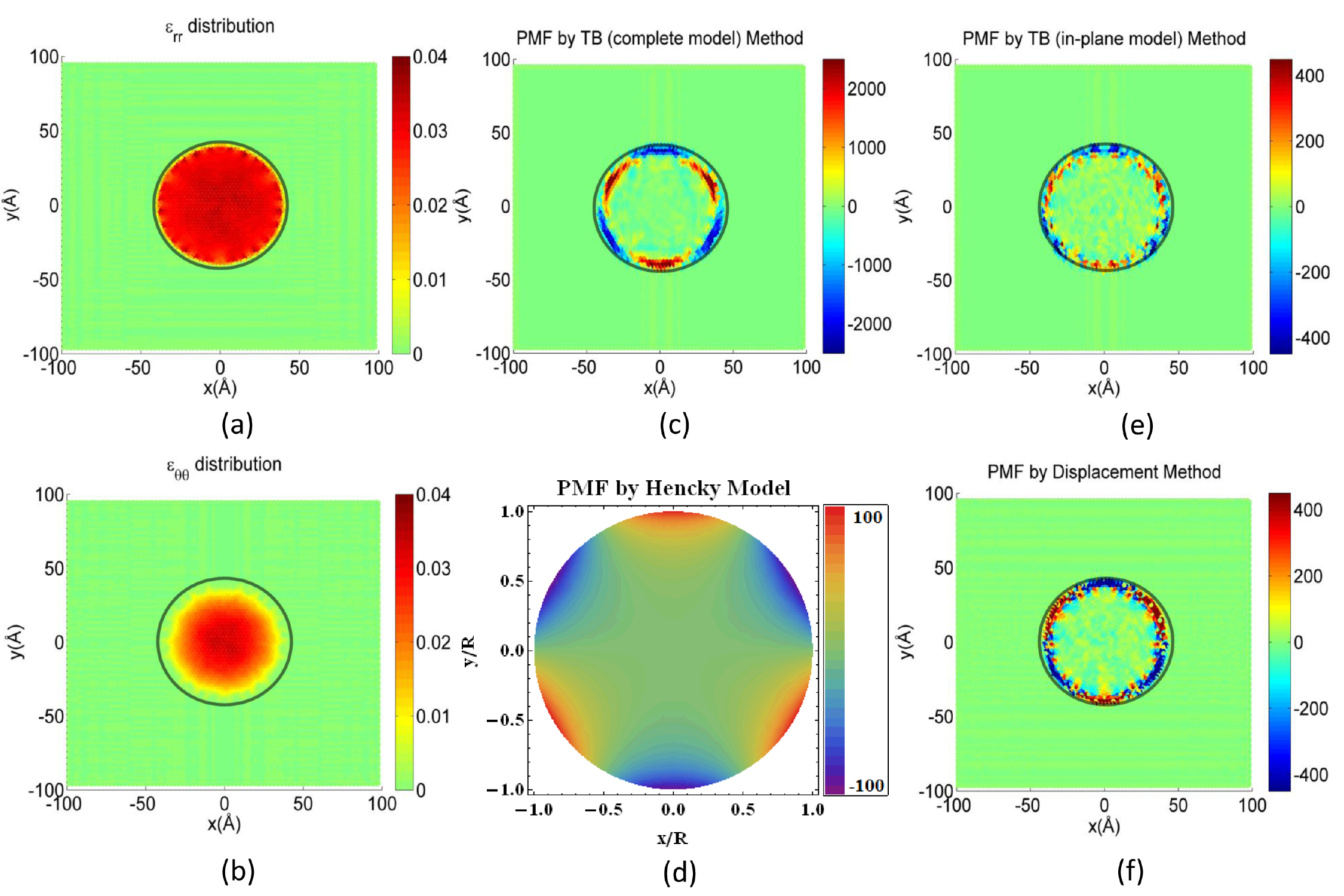}
\caption{(Color online) Results for a circular graphene bubble with 4\,nm 
radius and pressurized up to $\sim$1\,nm deflection; in this case graphene 
was clamped at the edge of the substrate aperture. (a) Radial strain, (b) 
tangential strain, (c) PMF by TB method with both in-plane and bending 
components, (d) PMF arising from Hencky's analytic model~\cite{FichterNASA1997} with 
the axes scaled in units of the circle radius, (e) PMF by TB method with 
in-plane component only, (f) PMF by displacement method. Note that, except for 
(d), all the panels refer to the same atomistic configuration. PMF in shown 
in units of Tesla. The edge of the substrate aperture used in the MD 
simulation is outlined (gray line) for reference.}
\label{fig:circle}\end{figure*}

We start with the most symmetric geometry, a circular graphene bubble, and 
compare the atomistic result with the continuum Hencky solution~\cite{FichterNASA1997}. 
In contrast to small deformation continuum models~\cite{KimPRB2011}, the Hencky 
model is valid for large in-plane (stretching) deformations, which lead to a 
different PMF distribution. To compute the PMFs associated 
with this analytical solution we used \Eqref{eq:Adef-strain}. 
Figs.~\ref{fig:circle}(c,d) show that the PMF distribution is dominated by very 
large magnitudes at the edges followed by a rapid decay towards the inside 
region of the nanobubble.  Both the MD and Hencky results show the 
six-fold symmetry expected for a cylindrically symmetric strain distribution; 
this agreement demonstrates the MD simulation successfully captures the strain 
distribution underlining the computed PMF. There are, however, two quite clear 
discrepancies between the PMF in these two figures: 
(i) Hencky's solution (panel d) yields values considerably smaller in magnitude 
than the calculation based on the MD deformations combined with 
Eqs.~\eqref{eq:Hopping} and \eqref{eq:Adef-full} (panel c); 
(ii) the sign of the PMF in panel (d) is \emph{apparently} 
reversed with respect to the sign of panel (c).
These discrepancies stem from the substantial bending present in graphene near 
the hole perimeter, and deserve a more detailed inspection in terms of the 
relative magnitude of the two contributions to the hopping variation: bond 
stretching and bond bending.

Since Hencky's result of \Fref{fig:circle}(d) hinges on \Eqref{eq:Adef-strain} 
that expresses the vector potential directly in terms of the strain tensor 
components, let us start by analyzing the predictions obtained by applying it to 
the atomistic case as well; to do that one computes the strain from the MD 
simulations using the displacement approach discussed earlier.
The result of that is shown in \Fref{fig:circle}(f), where the most important 
difference in comparison with \Fref{fig:circle}(c) is the significant reduction 
of the maximal fields obtained near and at the edges; this reflects the error 
incurred in the quantitative estimate of $B$ when the effect of bending is 
neglected. 
Note that, by construction, \Eqref{eq:Adef-strain} accounts only for the 
bond-stretching, and is accurate only to linear order in strain because it is 
based on a linear expansion of the hopping in the interatomic distances. 
Hence, in order to correctly extract from the atomistic simulations the total 
stretching contribution beyond linear order while still ignoring bending 
effects, we should calculate the PMF with the hopping as defined in 
\Eqref{eq:Hopping} (TB approach), but explicitly setting 
$\mbf{n}_i\cdot\mbf{n}_j=1$ and $\mbf{n}_i\cdot\mbf{d}=0$ (\ie assuming local 
flatness).
The outcome of this calculation is shown in \Fref{fig:circle}(e) which, in 
practical terms, is the counterpart of \Fref{fig:circle}(c) with 
bending effects artificially suppressed. In comparison with panel (f), it leads 
to slightly smaller PMF magnitudes. The linear expansion in strain of 
\Eqref{eq:Adef-full} thus slightly overestimates the 
field magnitudes, something expected because the hopping is exponentially 
sensitive to the interatomic distance and, by expanding linearly, one 
overestimates its rate of change with distance, overestimating the field 
magnitude as a result. One key message from \Fref{fig:circle} and the 
comparison between panel (c) and any of the subsequent ones is that the effects 
of curvature are significant at these scales of deflection and bubble size, 
particularly at the edge, where they clearly overwhelm the ``in-plane'' 
stretching contribution. We will revisit this in more detail in section 
\ref{sec:bending}.

%
\begin{figure} \begin{center} 
\includegraphics[width=0.45\textwidth]{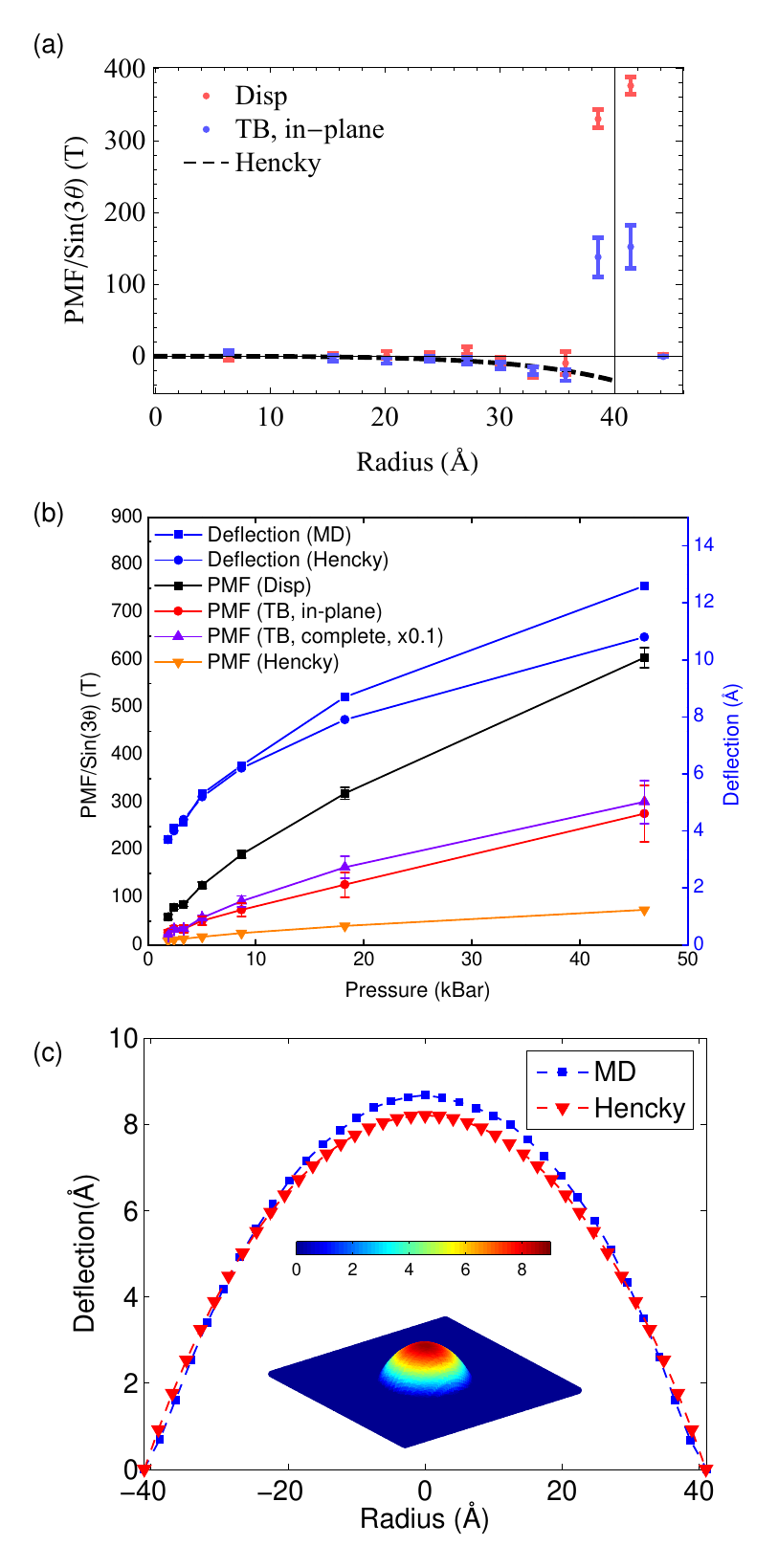}
\caption{(Color online) 
(a) Angular-averaged values of $B/\sin(3\theta)$ for the circular nanobubble 
with $R=4$\,nm considered in \Fref{fig:circle}. The different datasets 
correspond to different strategies discussed in the main text to obtain the 
PMF. The vertical line at $r=R\approx 40$\,\AA\ marks the radius of 
the circular aperture in the substrate. For $r<R$ the results extracted from MD 
closely follow the analytical curve, but there is a sharp sign change and 
increase at $r\approx R$ (see appendix \ref{ap:averaging} for details of the 
averaging procedure, as well as for the TB data including the full hopping 
perturbation).
(b) Comparison between the pressure-induced deflection and maximum PMF 
magnitude at the edge, $|\,B(R\approx 40\,\text{\AA})\,|$, obtained with the 
different approximations discussed in the text.
The points corresponding to the complete TB hopping are scaled by 0.1 for 
better visualization.
(c) A section of the simulated nanobubble (MD) at $\sim$19\,kBar and the 
corresponding Hencky's solution (the inset shows a 3D perspective of the former 
with the color scale reflecting the vertical displacement).
}
\label{fig:PBD} \end{center} \end{figure}

The second key message gleaned from \Fref{fig:circle} pertains to the 
importance of properly considering the boundary and loading conditions when 
analytically modeling the strain and deflection of graphene. This is related to 
the \emph{apparent} opposite sign in the PMF at the edge obtained from Hencky's 
solution in panel (d) when compared with all the other panels 
(containing the MD-derived results).
To elucidate the origin of the difference we show in \Fref{fig:PBD}(a) the PMF 
divided by the angular factor $\sin(3\theta)$, and averaged over all the angles 
(details discussed in appendix \ref{ap:averaging}). This plot provides a summary 
of the data in Figs.~\ref{fig:circle}(d,e,f) and allows a cross-sectional view 
of the variation of the field magnitude with distance from the center of the 
nanobubble. Direct inspection shows that the averaged MD data follows Hencky's 
prediction inside the bubble nearly all the way to the edge, at which point the 
PMF derived from the atomistic simulations swerves sharply upwards, changes 
sign, and returns rapidly to zero within one lattice spacing beyond the bubble 
edge (the curve derived from Hencky's model terminates at the edge, by 
construction). This effective sectional view explains why the density plots in 
Figs.~\ref{fig:circle}(c,d) seem to have an overall sign mismatch: in the 
MD-derived data, the plots of the PMF distribution are dominated by the 
large values at the edge which have an opposite sign to the field in the inner 
region. \Fref{fig:PBD}(a) shows that, rather than a discrepancy, there is a 
very good agreement between the strain field predicted by Hencky's solution and 
a fully atomistic simulation throughout most of the inner region of the 
nanobubble. However, since Hencky's solution assumes fixed boundary 
conditions at the edge (zero deflection, zero bending moment) 
\cite{FichterNASA1997}, it cannot capture the sharp bends expected at the atomic 
scale generated by the clamping imposed in these particular MD simulations (in 
effect, corresponding to zero deflection and its derivative). The finite 
bending stiffness of graphene \cite{Wei:2013} comes into play in that region, 
generating additional strain gradients which explain the profile and large 
magnitude of the PMF seen in the atomistic simulations.

In Fig.~\ref{fig:PBD}(b) we plot the evolution of the deflection and maximum 
PMF with increasing gas pressure. The maximum PMF is obtained around the edge 
of the aperture, and the values shown in the figure correspond to an angular 
average of the PMF amplitude there (see appendix \ref{ap:averaging} for 
details). The MD and analytical (Hencky's) solutions give comparable results for 
the deflection in the pressure range below $<1\times10^4$\,bar 
(\Fref{fig:PBD}(b), right vertical scale).  At higher pressures, 
Figs~\ref{fig:PBD}(b) and \ref{fig:PBD}(c) show that the analytical 
solution yields a slightly smaller deflection, as the underlying model 
does not capture the nonlinear elastic softening that has been observed in 
graphene in both experiments~\cite{leeSCIENCE2008} and previous MD 
simulations~\cite{junJN2011}. 
\Fref{fig:PBD}(b) includes also the maximum PMFs occurring at the bubble edge, 
when computed with the different approaches discussed above in connection with 
Figs~\ref{fig:circle}(c-f). We highlight that Hencky's solution cannot 
generate significant PMFs even at the largest deflections, whereas experiments 
in similarly sized and deflected nanobubbles easily reveal PMFs in the hundreds 
of Teslas~\cite{LevyScience2010,LuNatureComm2012}. This raises questions about 
the applicability of the Hencky solution at these small scales and large 
deflections.

The pressure required to rupture this graphene bubble was determined
to be around $1.9\times10^{5}$\,bar from our MD simulations.
Such a large value is required because of the small dimensions of
the bubble.
We can calculate the fracture stress by adopting a simple
model for a circular bulge test, \ie,  $\sigma \sim \frac{R\delta
P}{2w}$, where $\sigma$, $\delta P$, $R$, and $w$ are the stress,
pressure difference, radius, and thickness of the membrane,
respectively. Assuming $w$ to be 3.42\,\AA, we obtain a fracture strength of 
about 80\,GPa, which is in agreement with previous 
theoretical~\cite{zhaoJAP2010} and 
experimental~\cite{leeSCIENCE2008,zhaoJAP2010} results. Note that the plot in 
Fig.~\ref{fig:PBD} shows very large pressures (up to near the rupture limit of 
the bubble) and correspondingly large deflections since we 
wish to highlight the points of departure between the elastic model and the 
simulation results. Pressures and deflections considered in the specific cases 
discussed below are considerably smaller.

\begin{figure*} \centering
\includegraphics[width=0.85\textwidth]{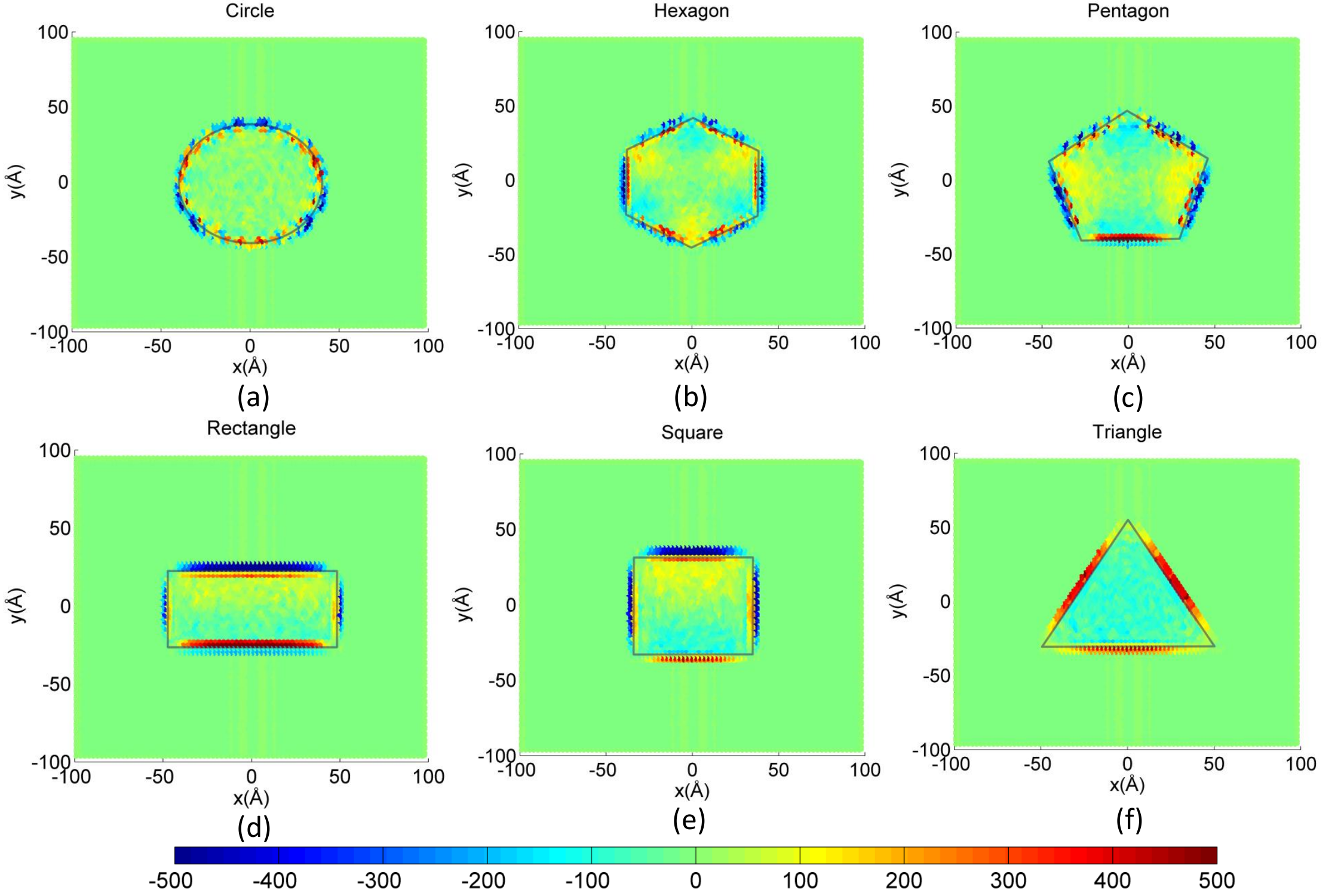}
\caption{
(Color online) Top views of PMF patterns for graphene bubbles of different 
geometries without substrate. (a) circle (b) hexagon (c) pentagon (d) rectangle 
(aspect ratio 1:2) (e) square (f) triangle. 
All the bubble areas are $\sim 50\,\text{nm}^2$, and side lengths and pressures 
can be found in the main text. In all cases, the graphene lattice is oriented 
with the zig-zag direction along the horizontal. The same color scale (in Tesla) 
is used in all panels. The edge of the substrate apertures used in the MD 
simulations is outlined (gray line) for reference.
}
\label{fig:allbubbles} \end{figure*}

With the good performance of the atomistic model on the circular graphene 
bubble established, we next extend the analysis to nanobubbles with different 
shapes. 
The bubbles are similarly obtained by inflation of graphene under gas pressure 
against a target hole in the substrate with the desired shape. 
Fig.~\ref{fig:allbubbles} shows results of a study of different shapes to which 
the displacement interpolation approach was applied to obtain the strain field 
and, thus, the PMFs.  The shapes are a square, a rectangle (aspect ratio of 
1:2), a pentagon, a hexagon, and a circle, and are presented in order of 
approximately decreasing symmetry. Those geometries are chosen because they are 
sufficiently simple that they can be readily fabricated experimentally with 
conventional etching techniques. The dimensions of the different bubbles were 
chosen such that their areas were approximately $\sim$50\,nm$^2$.  The pressure 
was 19000\,bar and side lengths for the bubble geometries shown in 
Figs.~\ref{fig:allbubbles}(a)-\ref{fig:allbubbles}(f) were, respectively, 4\,nm 
(circle), 4.4\,nm (hexagon), 5.7\,nm (pentagon), 5\,nm (rectangle, short edge), 
7.1\,nm (square), 10.6\,nm (triangle).

It is worth emphasizing that these features depend on 
the orientation of the graphene lattice with respect to the substrate 
aperture, as we would expect. This is clearly visible in the case of the square 
bubble in Fig.~\ref{fig:allbubbles}(e), for which the sharp magnetic field 
along the boundary is present along the horizontal (zig-zag) edges of the bubble 
but not along the vertical ones (armchair). This is also the reason why only 
the triangular aperture shown in Fig.~\ref{fig:allbubbles}(f) leads to a strong 
PMF that is nearly uniform as one goes around the boundary of the nanobubble.  
This is an important consideration for the prospect of engineering strained 
graphene nanostructures capable of guiding or confining electrons within, much 
like a quantum dot~\cite{qiNL2013}. The sharp PMF at the boundary acts 
effectively as a strong magnetic barrier, which might be tailored to confine 
some of the low energy electronic states 
\cite{Martino:2007,Vasilopoulos:2009,KlimovScience2012}.

The resulting PMF patterns in Fig.~\ref{fig:allbubbles} show that the highest 
values are found at the corners and edges of the different bubble shapes.  To 
illustrate more clearly the PMF patterns, we inflated the bubbles to large 
deflections ($\sim$1\,nm) with strains reaching 10\,\% and the corresponding 
pressure exceeding $1\times10^4$\,bar.  These large deflections explain why the 
PMF magnitudes in Fig.~\ref{fig:allbubbles} may reach over 500\,T. Given that 
the gas pressures used to achieve the results shown in this figure are rather 
high, some comments are in order.  

First, we emphasize that the relevant parameter is the deflection, rather than 
the pressure itself. In other words, gas pressure was employed here as 
\emph{one} way of generating graphene nanobubbles with predefined boundary 
geometries and target deflections, but other loading conditions might be used 
to achieve the same parameters. Our choice is motivated by the desire to 
constrain graphene and its interaction with the substrate as little as possible. 
Since we intend to reproduce bubbles with lateral size and deflections matching 
the magnitude of the values observed experimentally 
\cite{LevyScience2010,LuNatureComm2012} this requires large pressures (for a 
given target deflection $P$ is naturally smaller for larger apertures). 
Secondly, \citet{LuNatureComm2012} reported that experimental bond elongations, 
estimated from direct STM mapping of the atomic positions and deflections, can 
exceed 10\,\% in graphene nanobubbles on Ru. The high pressures considered in 
our MD simulations allow us to reach bond elongations of this order of 
magnitude.
Thirdly, pressures of the order of $10$\,kbar (1\,GPa) have been recently 
estimated to occur within nanobubbles of similar dimensions and deflections to 
the ones considered here, formed upon annealing of graphene-diamond interfaces 
\cite{CandyHaley}. Thus, pressures of this magnitude are not unrealistic in the 
context of nanoscale graphene blisters. 

\section{Substrate interaction: Graphene on Au\,(111)}\label{sec:Au}

Having considered the ideal case of graphene without a substrate, we move 
forward to study the more realistic case of graphene lying on an Au\,(111) 
substrate.  The main difference is that the carbon atoms are not rigidly 
attached to the substrate anymore outside the aperture, meaning that graphene 
can \emph{slide into the aperture} during inflation, subject to the 
interaction with the substrate.  This is an important qualitative difference, 
and reflects more closely the experimental situation, as recently reported 
in reference~\onlinecite{KittNL2013}.
The interatomic interactions were parameterized with 
$\epsilon_{C-Au}$=0.02936\,eV, $\sigma_{C-Au}$=2.9943\,\AA~\cite{PianaJPC2006}; 
$\epsilon_{C-Ar}$=0.0123\,eV, $\sigma_{C-Ar}$=3.573\,\AA~\cite{TuzunNano1996}; 
$\epsilon_{Ar-Ar}$=0.0123\,eV, 
$\sigma_{Ar-Ar}$=3.573\,\AA~\cite{RytkonenJCP1998}; the Ar-Au (gas-substrate) 
interactions were neglected to save computational time, and the substrate layer 
was held fixed for the entire simulation process.  Most of the graphene layer 
was unconstrained, except for a 0.5\,nm region around the outer edges of the 
simulation box where it remained pinned. Since the interaction with the 
substrate is explicitly taken into consideration, this approach realistically 
describes the sliding and sticking of graphene on the substrate as the gas 
pressure is increased, as well as details of the interaction with the substrate 
in and near the hole perimeter.

\begin{figure} \begin{center} 
\includegraphics[scale=0.4]{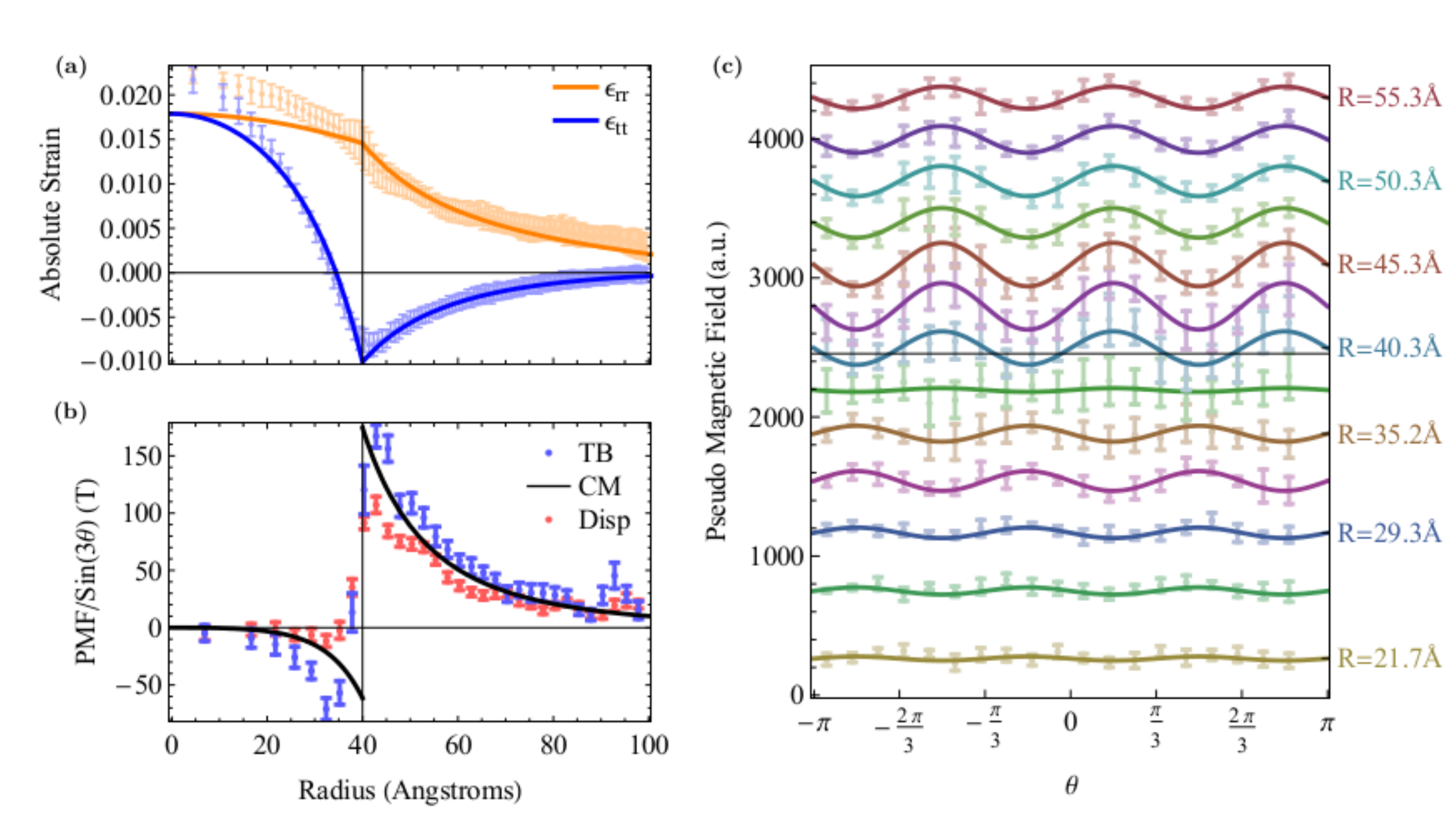}
\caption{(Color online)  (a) Strain components $\epsilon_{rr}$ and
$\epsilon_{\theta\theta}$ of a graphene bubble pressurized to a deflection of 
$\sim 1$\,nm against a circular hole with 4\,nm radius on a Au(111) substrate.
(b) The corresponding PMF along the radial direction from the bubble 
center computed according to the extended Hencky model~\cite{KittNL2013} (solid 
line) and from MD simulations within the TB (blue) or displacement (red) 
approach. Panel (d) shows the angular dependence of the PMF for selected radii.
}
\label{fig:circleAu} \end{center} \end{figure}

We start the discussion with a direct comparison of the deformation state of a 
circular bubble obtained from our simulations with the predictions of a 
recently developed and experimentally verified `extended-Hencky' 
model~\cite{KittNL2013} that accounts for the same sliding and friction effects. 
As can be seen in Fig.~\ref{fig:circleAu}(a), after fitting the friction in the 
continuum model to the MD simulation there is a very good agreement between the 
MD and extended Hencky results for the radial and tangential strains, 
$\epsilon_{rr}$ and $\epsilon_{\theta\theta}$, both in the inner and outer 
regions with respect to the substrate aperture.  The same good agreement is seen 
in the PMF profile extracted from the MD and analytical approaches, which is 
presented in Fig.~\ref{fig:circleAu}(b). The numerical data points shown in this 
panel represent an angular average over an annulus centered at different radii.
An important message from Fig.~\ref{fig:circleAu}(b) is that the maximum 
magnitude of the PMF occurs around the edge of the aperture, but \emph{on the 
outside} of the bubble. Whereas one expects the maximal PMFs to occur around the 
edge where the strain gradients are larger, the fact that the magnitude is 
considerably higher right outside rather than inside is not so obvious. This has 
important implications for the study of PMFs in graphene nanostructures but has 
been ignored by previous studies. It implies that models where only the 
deflection inside the aperture is considered (such as the simple Hencky model) 
can miss important quantitative and qualitative features.  They are captured 
here because the friction and sliding effects due to graphene-substrate 
interactions are naturally taken into account from the outset. One consequence 
is the ``leakage'' of strain outside the bubble region and the concurrent 
emergence of PMFs outside the aperture.  This should be an important 
consideration in designing nanoscale graphene devices with functionalities that 
rely on the local strain or PMF distribution.

\begin{figure*} \centering
\includegraphics[width=0.85\textwidth]{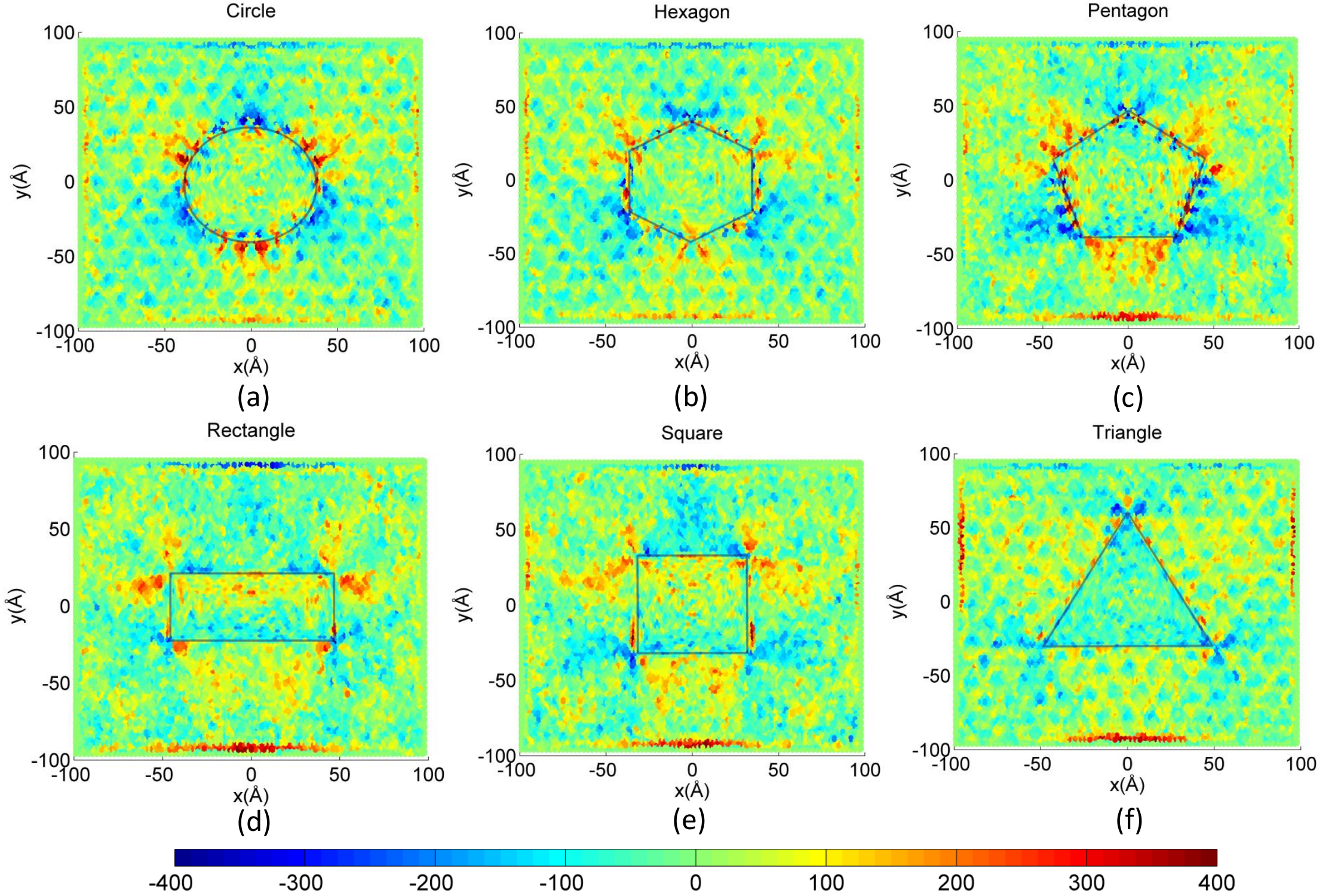}
\caption{
(Color online)  Top views of PMF patterns for graphene bubbles of different 
geometries on Au\,(111) substrates. 
(a) circle (b) hexagon (c) pentagon (d) rectangle (aspect ratio 1:2) (e) square
(f) triangle. All the bubble areas are $\sim 50\,\text{nm}^2$, and side
lengths and pressures can be found in the main text. In all cases, the 
graphene lattice is oriented with the zig-zag direction along the horizontal. 
The same color scale (in Tesla) is used in all panels. The edge of the substrate 
apertures used in the MD  simulations is outlined (gray line) for reference.
}
\label{fig:allbubblesAu} \end{figure*}

The other shapes studied on the Au\,(111) substrate are shown in 
Fig.~\ref{fig:allbubblesAu}.  The dimensions are the same as in 
Fig.~\ref{fig:allbubbles}, with an applied pressure of $\sim$\,30\,kbar. 
In addition to the appearance of non-negligible PMF outside the aperture 
region, a comparison with the data for bubbles clamped to the hole perimeter 
shows that now the PMF distribution inside is noticeably perturbed, and that 
the large field magnitudes observed in Fig.~\ref{fig:allbubbles} along the 
perimeter are considerably reduced and smoother.

\begin{figure} \centering
\includegraphics[scale=0.36]{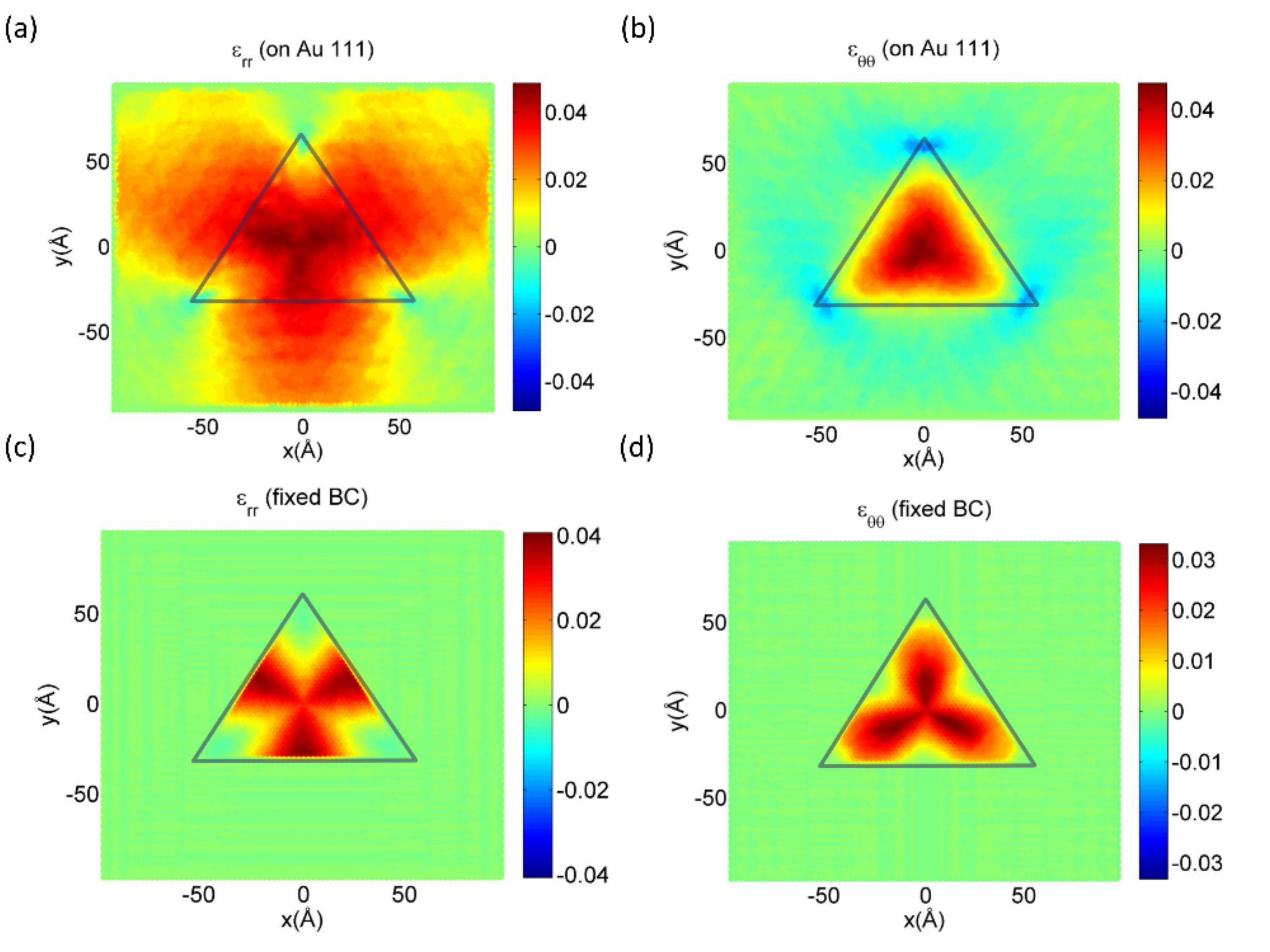}
\caption{(Color online)  
Spatial patterns of the strain tensor components $\epsilon_{rr}$ and 
$\epsilon_{\theta\theta}$ for a triangular bubble with a 10.6\,nm side. (a) 
and (b) pertain to graphene on a Au\,(111) substrate whose PMF profile has 
been shown in Fig.~\ref{fig:allbubbles}(f), while (c) and (d) 
correspond to the graphene bubble with an artificially fixed boundary 
condition whose PMF is shown in Fig.~\ref{fig:allbubblesAu}(f). The edge of the 
substrate aperture used in the MD simulation is outlined (gray line) for 
reference.
}
\label{fig:triangle} \end{figure}

To understand the origin of this difference, let us analyze in detail 
the representative case of a triangular nanobubble, as previous
experiments have shown that such nanobubbles can exhibit PMFs in excess
of 300\,T~\cite{LevyScience2010}.  Using our MD-based
simulation approach, we calculate the PMFs for triangular graphene
bubbles by inflating a graphene monolayer through a triangular hole in
the substrate. The set-up is as illustrated in Fig.~\ref{fig:schematic}, but 
with the circular hole replaced by a triangular one.
The triangular hole in the substrate had a side length of 10.6\,nm, and the
graphene sheet was inflated to a deflection of $\sim$\,1\,nm.
The resulting PMF distribution when one artificially clamps graphene outside 
the hole region has been shown in Fig.~\ref{fig:allbubbles}(f); the underlying 
strain components can be seen in Figs.~\ref{fig:triangle}(c,d). Upon 
inflation under the gas pressure, the geometry and the clamped conditions 
enforce an effective tri-axial stretching in the graphene surface that is 
clearly visible in the strain distribution. As pointed out by 
\citet{GuineaNaturePhy2010}, this tri-axial symmetry is crucial for the 
experimental observation of Landau levels in 
reference~\onlinecite{LevyScience2010} 
because it leads to a quasi-uniform PMF inside the nanobubble. Inspection of 
Fig.~\ref{fig:allbubbles}(f) confirms that the field is indeed of significant 
magnitude and roughly uniform within the bubble.

When the full interaction with the substrate is included and the graphene sheet 
is allowed to slip and slide towards the aperture under the inflation pressure, 
the geometry is no longer as effective as before in generating a clear triaxial 
symmetry: a comparison of the top and bottom rows of Fig.~\ref{fig:triangle} 
shows that the triaxial symmetry of the strain distribution is not so 
sharply defined in this case. Therefore, the finite and roughly uniform PMF 
inside the triangular boundary that is seen clearly in 
Figs.~\ref{fig:allbubbles}(f) [and also Figs.~\ref{fig:allbubbles-disp}(f)] is 
largely lost here.
To understand the difference we start by pointing out that the 
orientation of the triangular hole with respect 
to the crystallographic axes used here is already the optimum orientation in 
terms of PMF magnitude, with its edges perpendicular to the $\langle100\rangle$ 
directions (\ie, parallel to the zig-zag directions). 
Secondly, since the graphene sheet is allowed to slide, the strain distribution 
in the central region of the inflated bubble tends to be more isotropic, as we 
expect for an inflated membrane because of the out of plane displacement, and as 
can be seen in Fig.~\ref{fig:triangle}. 
This means that the trigonal symmetry imposed
on the overall strain distribution by the boundaries of the hole is
less pronounced near the center. As a result, even though strain
increases as one moves from the edge towards the center (as measured,
for example, by looking at the bond elongation directly from our MD
simulations), the magnitude of the PMF decreases because the trigonal
symmetry and strain gradients become increasingly less pronounced, and we
know that the isotropic (circular) hole yields zero PMF at the apex
(Fig.~\ref{fig:circle}).

The differences in trend and the sensitivity of the PMF distribution to 
the details of the interaction with the substrate highlight the importance of 
the latter in determining the final distribution and magnitude of the PMF, in 
addition to the loading, hole shape, and boundary conditions. In order to 
stress this aspect, and to make the role of the substrate interaction even more 
evident, we shall consider next a different metal surface.

\begin{figure*} \begin{center} 
\includegraphics[width=0.85\textwidth]{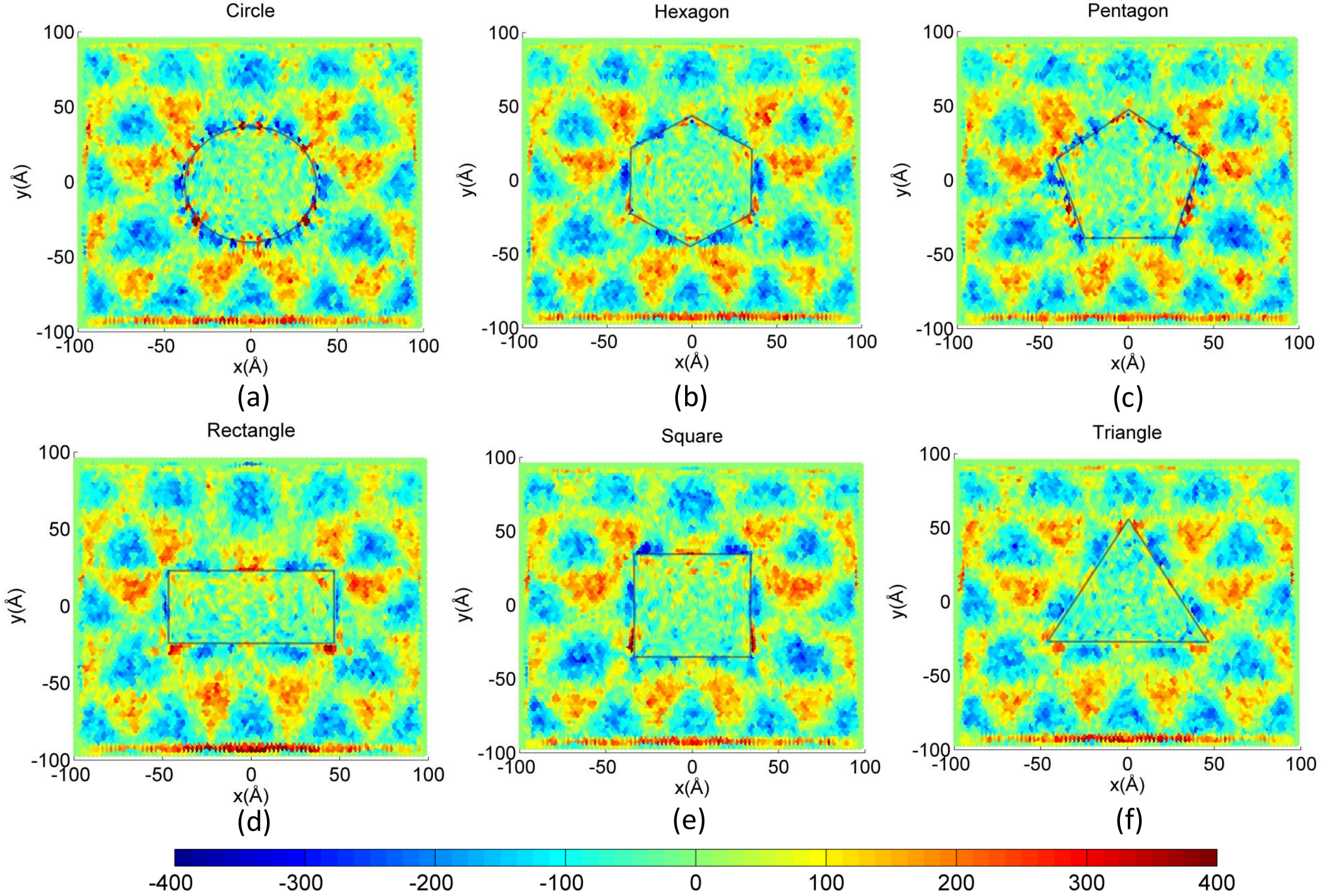}
\caption{(Color online)  Top views of PMF
patterns for graphene bubbles of different geometries on Cu\,(111) substrate. 
(a) circle
(b) hexagon (c) pentagon (d) rectangle (aspect ratio 1:2) (e) square
(f) triangle. All the bubble areas are $\sim 50\,\text{nm}^2$, and side
lengths and pressures can be found in the main text. In all cases, the 
graphene lattice is oriented with the zig-zag direction along the horizontal. 
The same color scale (in Tesla) is used in all panels. The edge of the substrate 
apertures used in the MD simulations is outlined (gray line) for reference.
} 
\label{fig:allbubblesCu} \end{center} \end{figure*}

\section{Substrate interaction: graphene on Cu\,(111)}\label{sec:Cu}

To gain further insight into the important effects of substrate interactions, we 
carried out simulations for a Cu\,(111) substrate, in addition to the Au\,(111) 
case considered above. This is in part motivated by a recent experimental 
study~\cite{heNL2012} showing that graphene grown by chemical vapor 
deposition on a Cu\,(111) substrate is under a nonuniform strain distribution. 
This nonuniform strain suggests that there might be interesting PMFs in the 
region of graphene surrounding the bubble. To analyze that we studied the PMF 
profile generated by the inflation of a graphene bubble constrained by a 
circular aperture with a radius of 4\,nm on a Cu(111) substrate. The Cu-C 
interactions were modeled using a Morse potential with parameters 
$D_0$=0.1\,eV, $\alpha$=1.7\,\AA, $r_0$=2.2\,\AA, and a cutoff radius of 6\,\AA 
\cite{MaekawaWEAR1995}.  Fig.~\ref{fig:allbubblesCu} shows the PMF 
distributions for differently shaped bubbles with deflection of $\sim$1\,nm on 
Cu\,(111) substrate. Despite the 
similarity between the geometry, dimensions, and deflections of this system and 
the one studied in Fig.~\ref{fig:allbubblesAu}, this one shows a much more 
pronounced 
modulation of PMF in the regions outside the aperture. In the same way 
that the Moir\'e patterns seen experimentally by \citet{heNL2012} reflect a 
non-negligible graphene-Cu interaction, the PMF distributions in 
Figs.~\ref{fig:allbubblesCu}(a-f) are much richer than in 
Figs.~\ref{fig:allbubblesAu}(a-f). That our simulation strategy 
involves pressing graphene against the substrate certainly 
enhances the interaction and promotes increased adhesion. This, in turn, adds a 
non-isotropic constraint for the longitudinal displacement and deformation of 
the graphene sheet which will affect the overall 
magnitude and spatial dependence of the PMF in the central region in such a way 
that, for this case, the PMF magnitude is higher outside the inflated portion 
of graphene, rather than inside or in the close vicinity of the boundary. This 
shows that the strain and PMF patterns in graphene can be \emph{strongly} 
influenced by the chemical nature of the substrate and not just its 
topography.

\begin{figure} \begin{center} 
\includegraphics[scale=0.35]{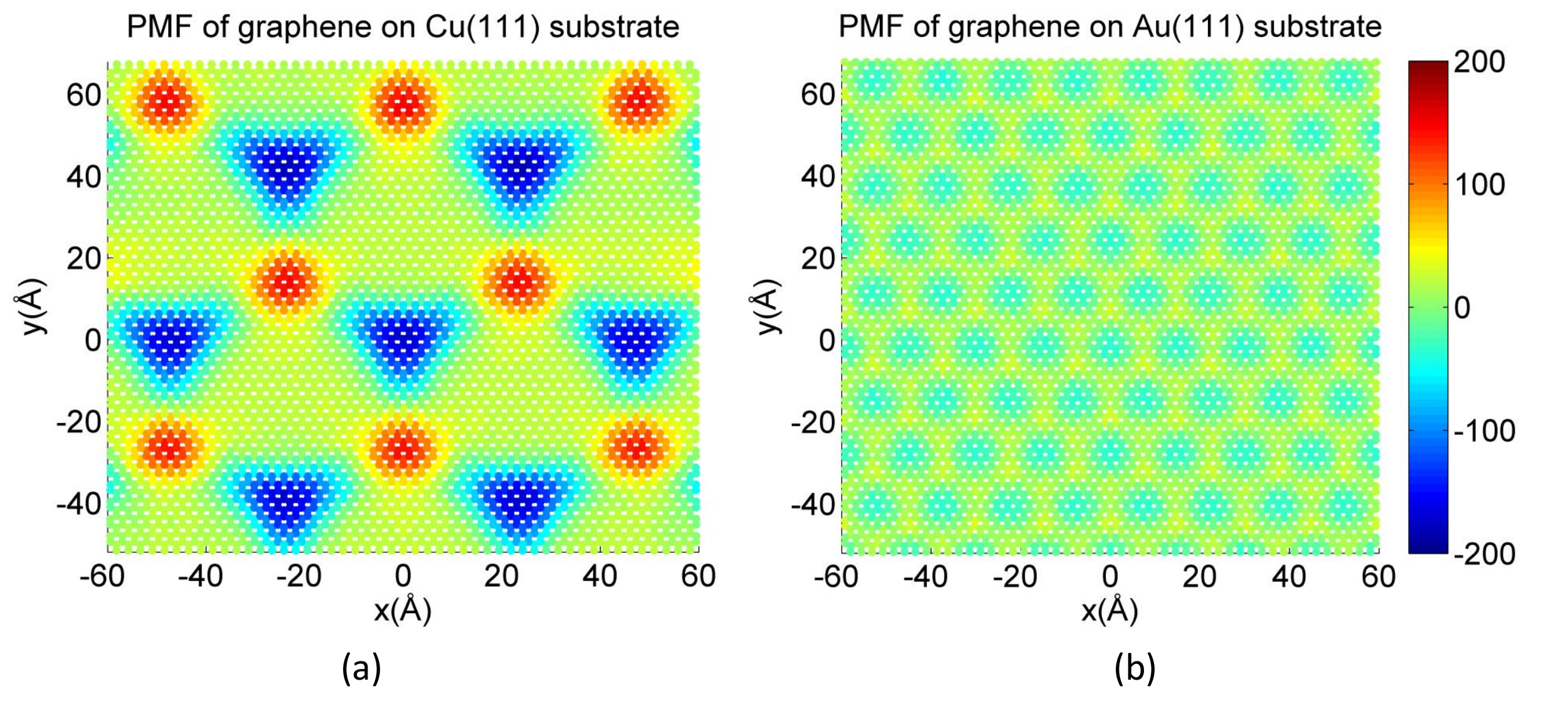}
\caption{(Color online)  PMF distributions of graphene on perfect (a) Cu\,(111) 
substrate and (b) Au\,(111) substrate without apertures nor gas pressure. The 
superlattice structure arises naturally from the need of the system to release 
strain buildup because of the mismatch in the lattice parameters of graphene 
and the underlying substrate. The PMF scale is in units of Tesla. 
} 
\label{fig:AuCu111} \end{center} \end{figure}

To reveal the PMF that is induced by the substrate alone, we show in 
Fig.~\ref{fig:AuCu111} a side-by-side comparison of the PMFs that result 
when graphene is let to relax on Au\,(111) and Cu\,(111), respectively. The 
plotted data were obtained from energy minimization without pressure or aperture 
to show the intrinsic effect of the two substrates.	
Several interesting features emerge from these results, the first of which 
being the spontaneous development of a superlattice structure with a 
characteristic and well defined periodicity that is different in the two 
substrates. This Moir\'e pattern in the PMF is the result of a corresponding 
 pattern in the strain field throughout the graphene sheet, which 
is caused by the need of the system to release strain buildup due to the 
mismatch in the lattice parameters of graphene and the substrate. 
A second important aspect is the considerable magnitude of the PMFs that 
can locally reach a few hundreds of Tesla just by letting graphene reach the 
minimum energy configuration in contact with the flat metal substrate. 
Another detail clearly illustrated by these two examples is the sensitivity to 
the details of the substrate interaction: the substrate-induced PMF on Cu can 
be many times larger than that on Au, and the Moir\'e period is also different. 
These super-periodicities are expected to perturb the intrinsic electronic 
structure of flat graphene whose electrons now feel the influence of this
additional periodic potential. That leads, for example, to the appearance of 
band gaps at the edges of the folded Brillouin zone. Such effects are currently 
a topic of interest in the context of transport and spectroscopic properties of 
graphene deposited on boron nitride, where this type of epitaxial strain 
is conjectured to play a crucial role in determining the metallic or insulator 
character  \cite{Woods:2014,SanJose:2014,NeekAmal:2014,Jung:2014}.

Since Fig.~\ref{fig:AuCu111} reveals a strong graphene-substrate interaction, 
it is not surprising that the PMF patterns in Fig.~\ref{fig:allbubblesCu} 
are still strongly dominated by the substrate-induced PMF. Unlike the 
cases discussed in \Fref{fig:allbubbles}, a significant structure remains in 
the PMF distribution outside the hole region due to the tendency of the lattice 
to relax towards the characteristic Moir\'e periodicity of \Fref{fig:AuCu111}(a) 
when in contact with a flat portion of substrate. 
In contrast, Au\,(111) has a larger lattice spacing and generates considerably 
less epitaxial strain in the graphene film, implying comparatively weaker PMFs. 
It is then natural that in the presence of the nanobubbles the geometry of the 
aperture dominates the final PMF distribution over the entire system when 
pressed against Au\,(111) (Fig.~\ref{fig:allbubblesAu}), whereas for Cu\,(111) 
the epitaxial contribution is the one that dominates 
(Fig.~\ref{fig:allbubblesCu}).

\section{Bending effects}\label{sec:bending}

The large deflection-to-linear dimension ratio in the inflated graphene bubbles 
analyzed so far calls for an analysis of the relative importance of the 
contribution to the PMF from bending in comparison with that from the local 
stretching of the distance between carbon atoms.

\begin{figure} \begin{center} 
\includegraphics[scale=0.36]{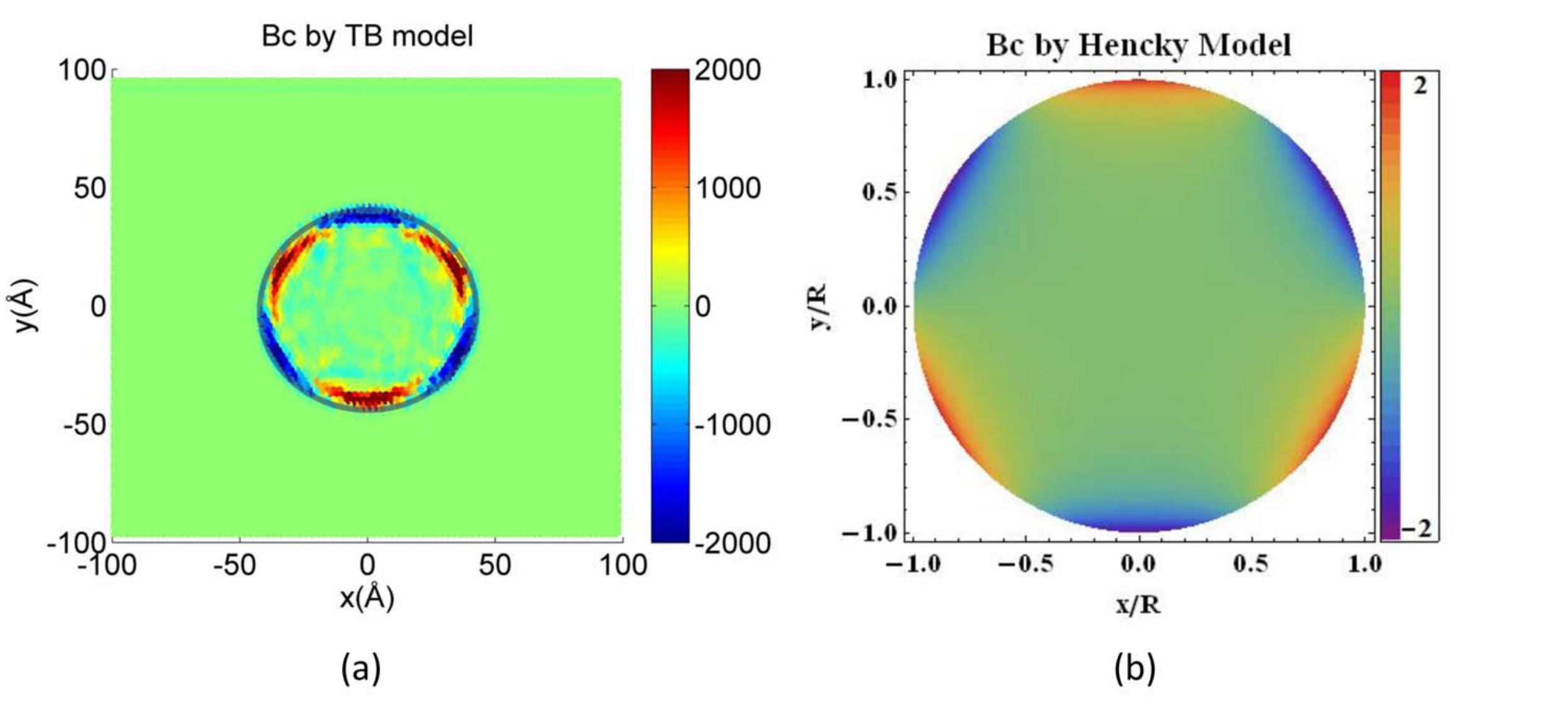}
\caption{(Color online) Density plot of the bending contribution to the 
pseudomagnetic field, $B^{(c)}$, for a circular graphene bubble with radius 
of 4\,nm and a deflection of $\sim 1$\,nm calculated by the TB method (a) and 
Hencky's model (b). The axes in (b) are scaled in units of the circle radius. 
The PMF scale is in units of Tesla. The edge of the substrate aperture used in 
the MD simulation is outlined (gray line) for reference.
} 
\label{fig:Bbc} \end{center} \end{figure}

When full account of stretching and bending is taken by replacing the hopping 
\eqref{eq:Hopping} in the definition of the vector potential $\mbf{A}$ given 
in \Eqref{eq:Adef-full} the resulting PMF can have considerably higher 
magnitudes, as was already seen in \Fref{fig:circle}(f). To isolate the effect 
of bending alone one can split the full hopping \eqref{eq:Hopping} in two 
contributions, $t_{ij} = t_{ij}^{(xy)} + t_{ij}^{(c)}$, where the 
``in plane'' stretching term is simply
\begin{equation}
  - t_{ij}^{(xy)} = V_{pp\pi}(d).
\end{equation}
Since the gauge field $\mbf{A}$ is a linear function of the hopping 
\eqref{eq:Adef-full}, it can be likewise split into the respective 
stretching and bending contributions so that $\mbf{A} = \mbf{A}^{\!(xy)} + 
\mbf{A}^{\!(c)}$. When the PMF associated with $\mbf{A}^{\!(c)}$ is thus 
calculated for the circular bubble of \Fref{fig:circle} we obtain the result 
shown in \Fref{fig:Bbc}(a). As was already seen when comparing the different 
PMF curves in \Fref{fig:PBD}, the effect of the curvature at the edges is quite 
remarkable and overwhelmingly dominant in that region.

\begin{figure} \centering
\includegraphics[width=0.45\textwidth]{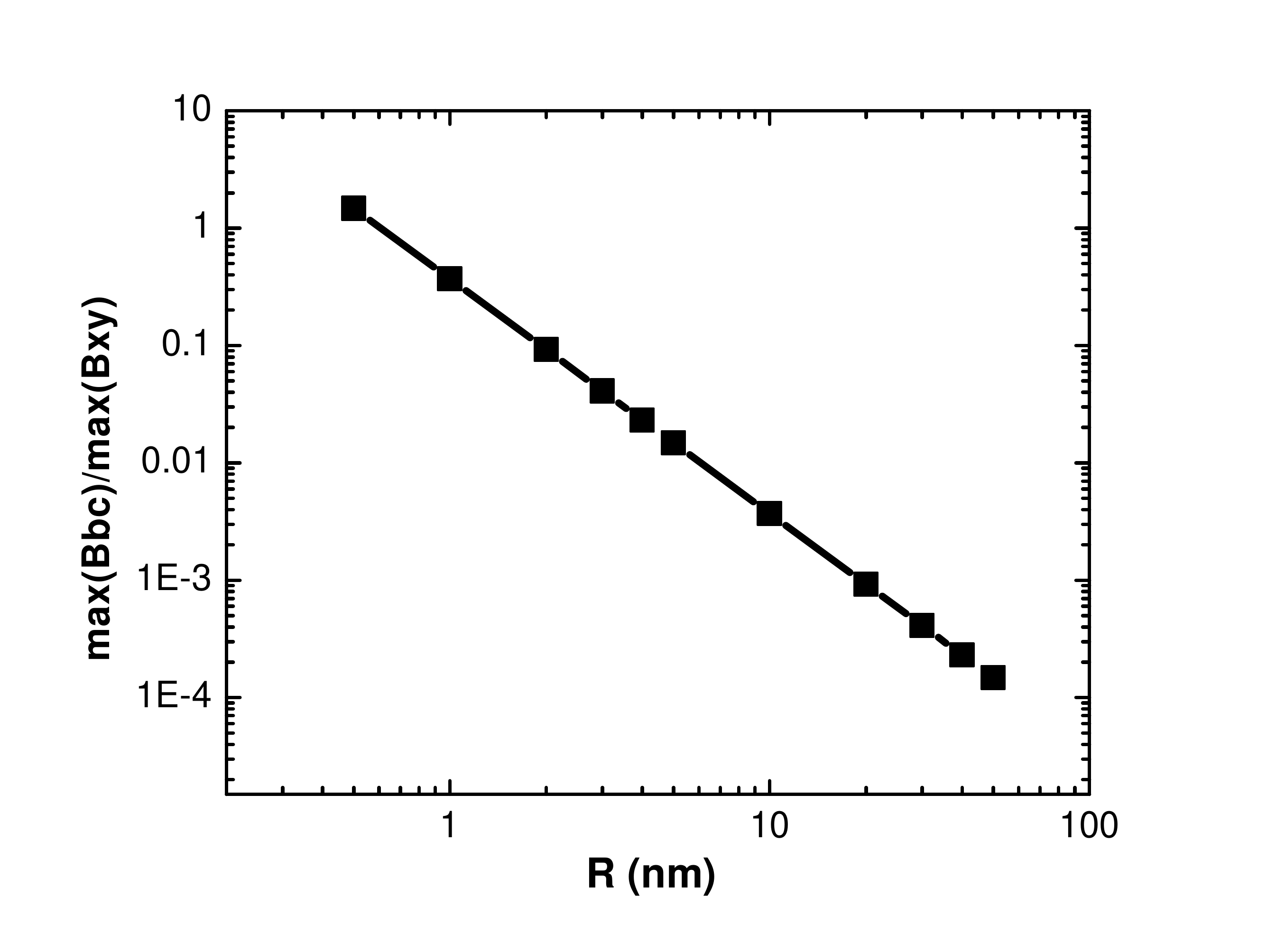}
\caption{(Color online)  Ratio of the maximum PMF induced by bending and 
stretching ($B_c/B_{xy}$) for circular graphene bubbles as a
function of the graphene radius $R$, according to Hencky's solution.
} 
\label{fig:BbcBxy}\end{figure}

More importantly, this fact could have been underappreciated if the stretching 
and bending contributions had been extracted only on the basis of an analytical 
solution of the elastic problem such as Hencky's model. To be definite in this 
regard let us consider the magnitude of the contribution to the PMF that comes 
from bending in the continuum limit. If a gradient expansion of the full 
hopping \eqref{eq:Hopping} is performed, the vector potential 
\eqref{eq:Adef-full} can be expressed in terms of quadratic combinations of the 
second derivatives of the deflection $h(x,y)$ \cite{PereiraPRL2010}. For 
example, the term $V_{pp\pi}(d)\, \mbf{n}_i\cdot\mbf{n}_j$ in \eqref{eq:Hopping} 
leads to 
\begin{subequations} \label{eq:Adef-bending}
\begin{align}
  A_{x}^{\!(c)} &= - \frac{3 a^2 V_{pp\pi}^0}{8qv_F} 
\left[\Bigl(\frac{\partial^2 {h}}{\partial {y^2}}\Bigr)^{\!\!2} -
\Bigl(\frac{\partial^2 {h}}{\partial {x^2}}\Bigr)^{\!\!2}\right]\!\!, \\
  A_{y}^{\!(c)} &= - \frac{3 a^2 V_{pp\pi}^0}{4qv_F} 
\left[\frac{\partial^2{h}}{\partial{x}\partial{y}}
\Bigl(\frac{\partial^2 {h}}{\partial {y^2}} + \frac{\partial^2
{h}}{\partial {x^2}}\Bigr)\right]\!\!.
\end{align}
\end{subequations}
This particular contribution was previously discussed by~\citet{KimEPL2008} 
and, since all the bending terms have the same scaling $\sim a^2 h^2/R^4$, where 
$h$ and $R$ are the characteristic height and radius, respectively, 
consideration of this one alone suffices for our purpose of establishing the 
magnitude of the bending terms in comparison with the stretching one. Replacing 
the deflection $h(x,y)$ provided by Hencky's solution in 
Eqs.~\eqref{eq:Adef-bending} leads to the result shown 
in \Fref{fig:Bbc}(b); it is clear that the maximum $B_c$ so obtained at 
the edges is much smaller than the one derived from the atomistic simulation 
with the full hopping. It is not surprising that the PMF coming from bending at 
the level of Hencky's model is so small.  A simple scaling 
analysis of the vector potentials in the continuum limit shows that, from 
\Eqref{eq:Adef-strain}, $\mbf{A}_{xy}$ scales with strain as 
$\mbf{A}_{xy}\sim\epsilon$ and strain itself scales with deflection as 
$\epsilon\sim(h/R)^2$ for a characteristic linear dimension $R$ of the 
bubble. On the other hand, from \eqref{eq:Adef-bending} $\mbf{A}_c$ scales like 
$\mbf{A}_c\sim(a h)^2/R^4$. Therefore, the ratio $B_c/B_{xy}$ will scale 
as $\sim (a/R)^2$. Since the bubble under analysis has $a/R \approx 0.04$ the 
bending contribution is indeed expected to be much smaller than the stretching 
one. We can even be more quantitative and extract the maximum values of 
$B_c$ and $B_{xy}$ from Hencky's solution and compare their relative 
magnitudes as a function of circle radius, as shown in Fig.~\ref{fig:BbcBxy}. 
Hencky's solution predicts that only when the radius of the circular bubble 
decreases below about 1\,nm does the contribution of the curvature-induced 
pseudomagnetic field become of the same order as that due to in-plane 
stretching. This situation is equivalent to the need to account for 
the curvature and orbital re-hybridization when describing the electronic 
structure of carbon nanotubes with diameters below length scales of this same 
magnitude at the tight-binding level \cite{Blase:1994,Kane:1997}; the neglect of 
these effects in the nanotube case leads to incorrect estimation of the band 
gaps and even of their metallic or insulating character. 

The problem with these considerations is that they fail to anticipate 
the large effect at the edges, particularly the scaling analysis which tells 
us only about the relative magnitude of bending \emph{vs} stretching in the 
central region. But, because we are inflating graphene under very high 
pressures in order to achieve deflections of the order of 1\,nm, a sharp bend 
results at the edge of the substrate aperture through which graphene 
can bulge outwards; it is this curvature effect that dominates the PMF plot in 
\Fref{fig:Bbc}, not the overall curvature of the bubble on the large scale. 
Hencky's solution cannot capture this since it is built assuming zero 
radial bending moment at the edge \cite{FichterNASA1997}. Moreover, since this 
happens within a distance of the order of the lattice constant itself, the 
details of the displacements at the atomistic level including non-linearity and 
softening at large strains and curvatures become crucial. This further 
highlights the importance of accurate atomistic descriptions of the deformation 
fields in small structures such as the sub-5\,nm graphene bubbles we have 
considered in this paper, and which have been shown experimentally to lead to 
significant pseudomagnetic fields~\cite{LevyScience2010,LuNatureComm2012}; at 
this level models based on continuum elasticity theory can become increasingly 
limited for accurate quantitative predictions and should be applied with 
caution.

\begin{figure} \centering
\includegraphics[width=0.5\textwidth]{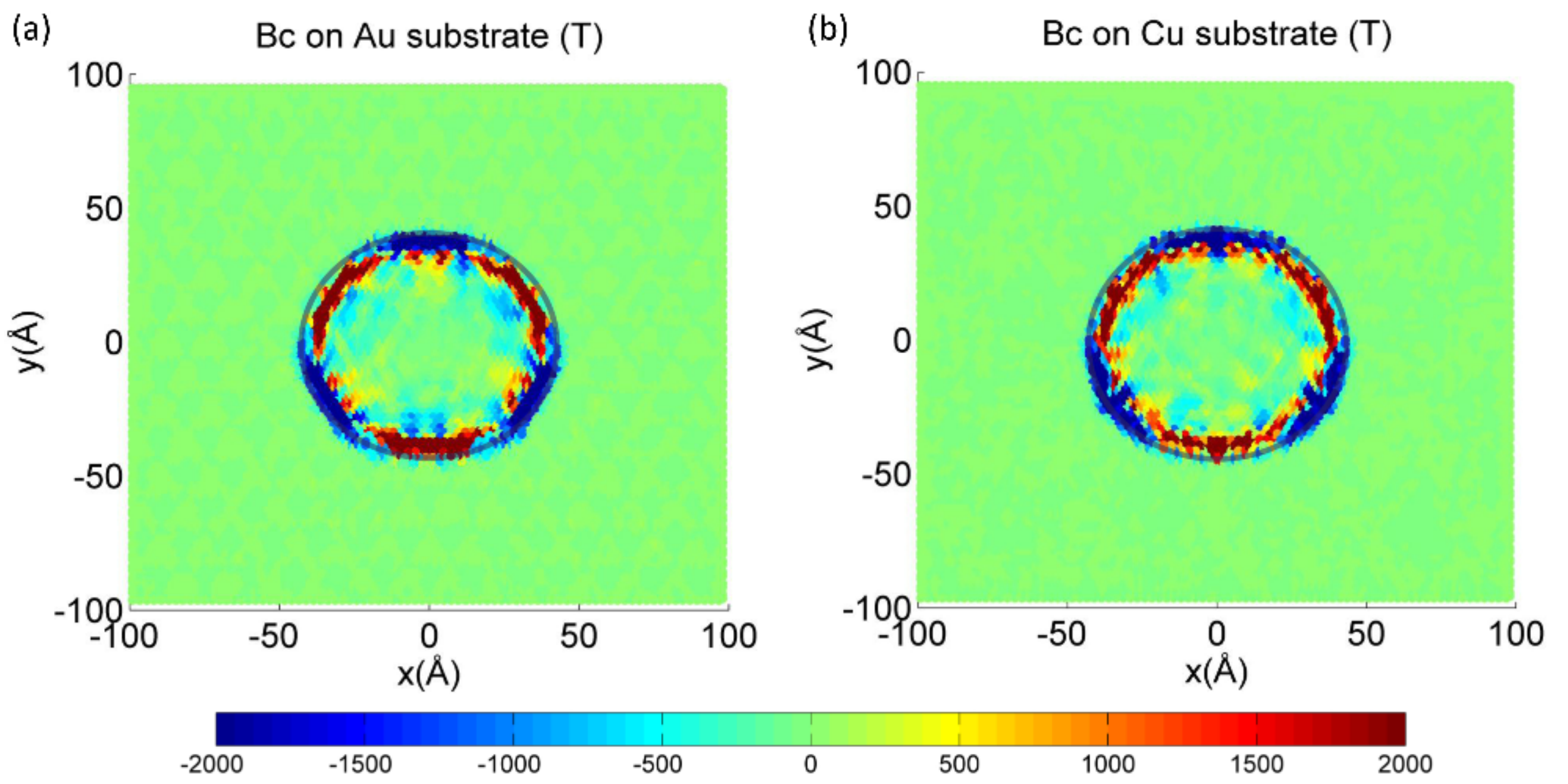}
\caption{(Color online)  Density plot of the bending contribution to the 
pseudomagnetic field, $B^{(c)}$, for a graphene bubble deflected to $\sim 
1$\,nm upon pressuring through a circular aperture of radius 
4\,nm in a Au (a) and a Cu (b) substrate. The PMF scale is in units of Tesla.
The edge of the substrate apertures used in the MD simulations is outlined (gray 
line) for reference.
} 
\label{fig:BbcAuCu}\end{figure} 

Finally, when realistic substrate conditions are considered, one can see that 
the slippage effects contribute very differently for the PMFs arising from 
stretching and from bending. A general feature of the PMF distribution obtained 
with realistic Au and Cu substrates is its smaller overall magnitude in 
comparison with the artificially clamped nanobubbles. This is easy to 
understand because the ability to slide in contact with the substrate 
allows graphene to stretch not only in the bubble region, but essentially 
everywhere, thereby reducing the strain concentration around the edge of 
the aperture; and with smaller strain gradients one gets smaller PMFs. The 
bending effects, on the other hand, are not expected to be much affected by the 
sliding, especially when comparing nanobubbles with the same amount of 
vertical deflection, because the sharpness of the bend at the edge of the 
aperture is constrained mostly by the geometry alone. Direct inspection of the 
contribution to the PMF arising from curvature in the Au and Cu substrates 
directly confirms this intuitive expectation, as shown in \Fref{fig:BbcAuCu}. 
Just as in the clamped case where graphene is pinned to the substrate and cannot 
slide, the PMF associated with bending is seen to dominate the field 
distribution, with magnitudes similar to the registered in \Fref{fig:Bbc}, 
and much larger than the PMF in the center of the bubble or the substrate 
region (cf. Figs.~\ref{fig:allbubblesAu} and \ref{fig:allbubblesCu}). This not 
only shows how crucial the PMF associated with bending can be in certain 
approaches to generate graphene nanobubbles, but also that it is an effect 
largely insensitive to the details of the substrate.

\section{Discussion and conclusions}

We have evaluated the strain-induced pseudomagnetic fields in pressure-inflated 
graphene nanobubbles of different geometries and on different substrates whose 
configurations under pressure were obtained by classical MD simulations.
The geometry of the nanobubbles is established by an aperture of prescribed 
shape in the substrate against which a graphene monolayer is pressed under gas 
pressure. Our results provide new insights into the nature of pseudomagnetic 
fields determined by the interplay of the bubble shape and the degree of 
interaction with the underlying substrate. 
On a technical level, if bending is (or can be) neglected, we have established 
that an approximate {\it displacement-based} approach is adequate to obtain the 
strain tensor and accurate values of the pseudomagnetic fields from MD 
simulations when compared with a direct {\it tight-binding} approach where the 
modified hoppings are considered explicitly. 

By comparing nanobubbles inflated in three different substrate scenarios --- 
namely, an arguably artificial, simply clamped graphene sheet with no substrate 
coupling and more realistic conditions where the full interaction with
Au\,(111) and Cu\,(111) substrates is included from the outset in the MD 
simulations --- we demonstrated that the graphene-substrate interaction is an 
essential aspect in determining the overall distribution and magnitude of 
strain and the PMFs both inside and outside the aperture region. For example, 
sections \ref{sec:Au} and \ref{sec:Cu} demonstrate that graphene can adhere 
substantially to the substrate in atomically flat regions leading to sizable 
PMFs stemming only from epitaxial strain, even in the absence of any pressure 
or substrate patterning. This adhesion varies from substrate to substrate and, 
in the presence of an aperture or other substrate patterning, perturbs the 
final strain distribution of the nanobubble when compared with a simply clamped 
edge. On a more quantitative level, in the cases analyzed here where the 
aspect ratio of the bubbles is close to 1, the magnitude of the PMFs 
associated with epitaxial strain alone can easily be of the same magnitude as 
the PMF generated within the bubble region. For Cu this is clear in 
Figs.~\ref{fig:allbubblesCu} and \ref{fig:AuCu111}, and implies that the 
presence of the aperture is not the main factor determining the field 
distribution. 

\begin{figure} \centering
\includegraphics[width=0.4\textwidth]{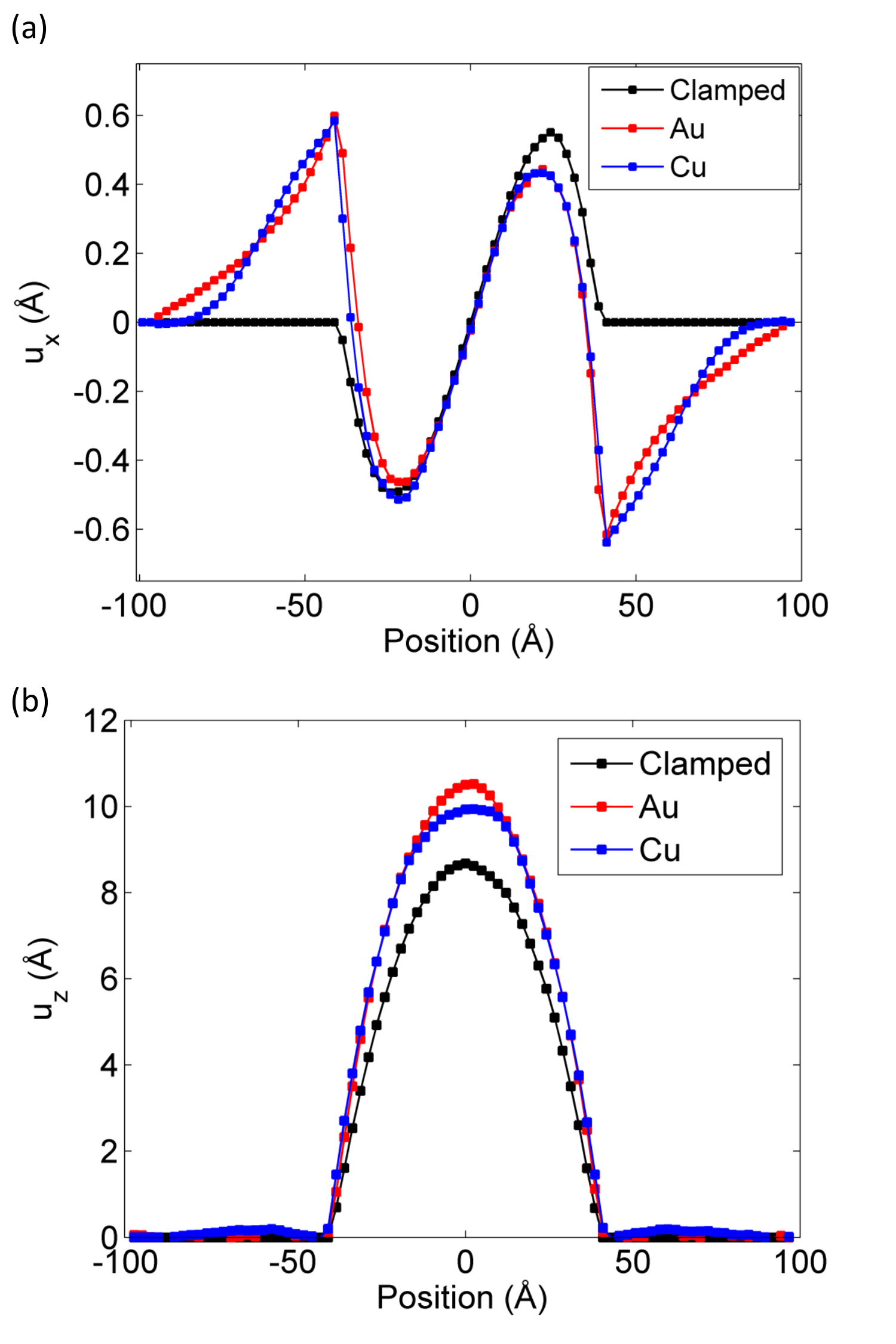}  
\caption{(Color online) The radial and vertical displacement components of 
graphene along the diametral section $y=0$ of three circular nanobubbles of 
radius 4\,nm, corresponding to the three substrate conditions considered in 
this report. Panel (a) shows the component $u_x$ of the graphene in-plane 
displacement which corresponds, in essence, to the radial displacement 
because of the circular symmetry. Panel (b) shows the vertical deflection. All 
cases were inflated under the same pressure of $\approx 19$\,kBar.
} 
\label{fig:slippage}\end{figure}

To better appreciate this aspect, we can inspect the averaged cross-section of 
the PMF provided in \Fref{fig:circle_sectional} whose the details are given in 
Appendix~\ref{ap:sectional-PMF-circular}. The key message conveyed by the data 
there is that under more realistic conditions describing the 
graphene-substrate interaction, and for the range of parameters explored here, 
the PMFs are no longer concentrated in and around the aperture. The section 
shown reveals that the PMF can be considerably higher in regions well outside 
the aperture than inside or near the edge. This arises because, on the one hand,
the graphene-substrate interaction alone is able to generate considerable local 
strain gradients that beget PMFs as large as the ones that appear by 
forcing the inflation of graphene through the aperture [cf. 
\Fref{fig:AuCu111}]. On the other hand, the fact that graphene can slip into 
the aperture when simulated on the Au and Cu substrates softens the strain 
gradients in its vicinity in comparison with the artificially clamped 
scenario. Slippage under these realistic substrate conditions prevents strain 
from concentrating solely within the aperture region which, instead, spreads to 
distances significantly away from the aperture. This is a sensible outcome on 
account of the very large stretching modulus of graphene that tends to penalize 
stretching as much as possible. We illustrate this behavior in 
\Fref{fig:slippage}(a) that compares the magnitude of the radial displacement 
of circular nanobubbles. Whereas in the artificially clamped nanobubble 
graphene remains undisturbed (by design) outside the aperture, it is clear that 
in either the Au or Cu substrates the carbon atoms pertaining to the region 
initially outside the aperture are radially pulled everywhere towards it under 
pressure, as one intuitively expects. One consequence of this is the softening 
of strain gradients in the bubble region: slippage naturally tends to diminish 
the PMFs generated within and around the aperture. The other is that, 
obviously, the deflection at the center is increased, as shown in 
\Fref{fig:slippage}(b).

The joint effect of these two factors (slippage and adhesion to the substrate) 
is that forcing graphene into a nanobubble profile at the center of the system 
is no longer effective in concentrating the strain gradients and, consequently, 
the PMF is no longer more prominent there. One immediate implication of this is 
the fact that whether or not it is feasible to locally tailor the PMF 
distribution on very small (nanometer) scales depends not only on the elastic 
response of graphene or its loading and geometric constraints, but also on the 
nature of the substrate involved.

It is also clear from the above that there might exist certain substrates in 
which the epitaxial strain can be significant enough to, by itself, lead to 
visible modifications of the electronic structure of graphene, and even lead to 
modified transport characteristics \cite{SanJose:2014,Jung:2014,NeekAmal:2014}. 
Incidentally our pressure-based approach facilitates and promotes a uniform 
adhesion because graphene is compressed against the substrate. It would 
be interesting to experimentally study graphene on top of such substrates 
inside pressure chambers, and assess the degree of control that can be 
achieved over the Moire patterns and the modifications of the electronic and 
transport characteristics. 

Another important factor to consider in estimating the magnitude and profile of 
the PMF generated under a given set of force distributions and geometric 
constraints is whether those conditions lead to strong local curvature of 
the graphene lattice. We analyzed this issue here by separately considering the 
contributions from bond bending and from stretching to the PMF in the 
representative case of circular nanobubbles. Our results establish that, even 
though the overall, large-scale curvature of the graphene sheet leads 
to significant corrections to the pseudomagnetic field only in ultra-small 
bubbles with diameter smaller than 2\,nm, sharp bends arising from direct 
clamping or from being pressed against an edge in the substrate aperture result 
in much stronger PMFs locally. At the qualitative level this is 
naturally expected and certainly not surprising. What is surprising significant 
is that the bending contribution can be many times larger than its stretching 
counterpart, leading to a PMF distribution dominated by large values near the 
edges of the substrate apertures. Moreover, since this is a local 
geometric effect, it does not depend on the bubble size but only on the local 
curvature around sharp bends, and should remain in considerably larger systems. 
This indicates that curvature of the graphene sheet should certainly not be 
ignored in many situations involving out-of-plane deflection, even though the 
scaling analysis based on the overall profile could point otherwise.

Finally, we underline once more that the strategy to generate graphene 
nanobulges through gas pressure was chosen here to minimize other external 
forces and influences on the deflection and slippage of graphene while being 
able to produce deflections and aspect ratios equal to those reported 
in recent experiments that explore the local electronic properties of these 
structures. But the conclusions and implications discussed above certainly carry 
to various other means of achieving such or similar nanobubbles and have, 
therefore, a wide reach and wide import beyond graphene pressurized through 
apertures.

\textit{Note added}. Recently, we became aware of a recent proposal  to connect structure and electronic properties of two-dimensional crystals based on concepts from discrete geometry that allows yet another efficient alternative to obtain the strain and PMF at discrete lattice points without the need, for example, to perform numerical derivatives upon the displacement fields or vector potentials extracted from the MD data \cite{BarrazaLopez:2013,Sanjuan:2014}.

\acknowledgments

ZQ acknowledges support from a Boston University Dean's Catalyst Award and the 
Mechanical Engineering Department at Boston University.  ZQ and HSP acknowledge 
support of U.S. National Science Foundation grant CMMI-1036460. The authors thank Prof. Teng Li 
for illuminating discussions on the stress method. VMP and AHCN are supported 
by the NRF CRP grant ``Novel 2D materials with tailored properties: beyond 
graphene'' (R-144-000-295-281). This work was supported in part by the U.S. 
National Science Foundation under grant No. PHYS-1066293 and the hospitality of 
the Aspen Center for Physics.

%
%

\appendix

\section{Angular averaging of the PMF}%
\label{ap:averaging}

\begin{figure} \centering
\includegraphics[width=0.45\textwidth]{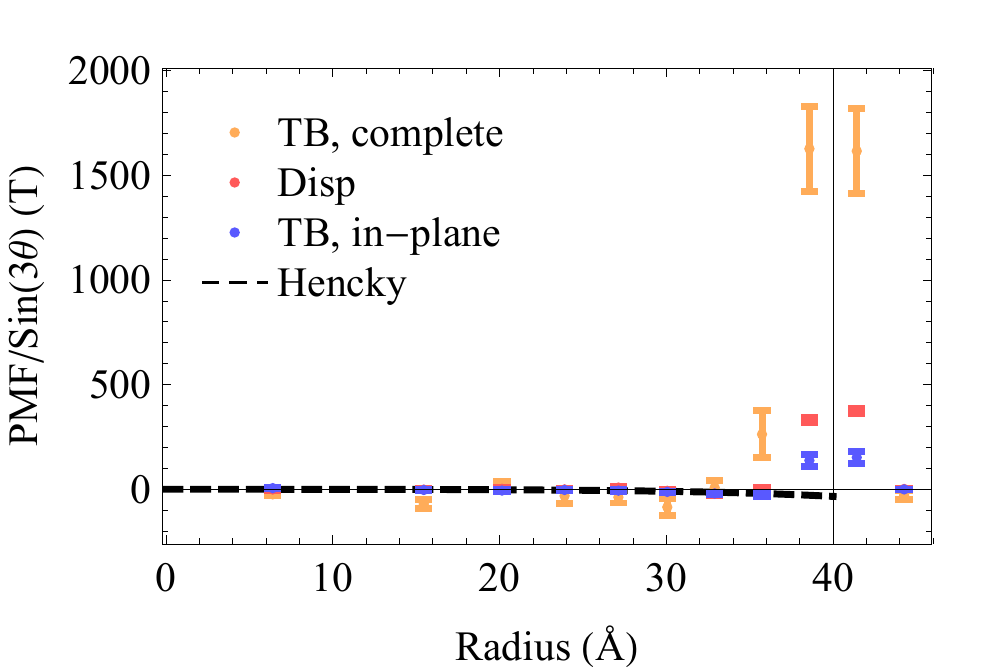}
\caption{(Color online) Angular-averaged amplitude of the PMF for the same 
cases presented in \Fref{fig:circle} in the form of density plots. The 
horizontal axis represents the distance from the center of a pressurized 
circular graphene nanobubble with clamped boundary conditions. The data 
contained here is the same shown in \Fref{fig:PBD}(a), except that here the 
(orange) data corresponding to the PMF obtained from the full 
hopping perturbation [\Eqref{eq:Hopping}] is included for comparison as well. 
The bending effects are clearly dominant around the edge/clamping region. Away 
from the edge, and inside, the three numerical curves follow Hencky's model.
}
\label{fig:circle_angular}\end{figure}

\Fref{fig:circle_angular} below shows the radial dependence of the averaged PMF 
amplitude close to the edge of the circular aperture for the clamped circular 
case discussed in section \ref{sec:clamped} (\Fref{fig:circle}).
In \Fref{fig:PBD} we plot the amplitude of the PMF at the edge of the circular 
aperture in the substrate for various inflation pressures with clamped 
graphene. In \Fref{fig:circleAu}(b) we show the average amplitude of the PMF at 
different distances from the center. 

In all these cases, the data shown reflect the PMF amplitude averaged over the 
azimuthal direction. To extract the average PMF at a given radius, the 2D 
distribution of the field is divided into a sequence of radial and azimuthal 
bins (annular sectors). For each radial annulus there are 20 bins, each with a 
18 degree width. The width of the radial annulus is chosen such that at least 10 
atoms lie in each bin (this is why there are fewer data points near the center 
of the bubble). The average and standard deviations of the PMF in each bin 
correspond to the value and 
error bar of that bin. For example, each point in \Fref{fig:circleAu}(c) 
corresponds to this average PMF for a given bin. Afterwards, for each radial 
annulus the data is fit to the expected $\sin(3 \theta)$ dependence. The 
amplitude of the best fit is plotted as a point [e.g., as 
in \Fref{fig:circle_angular}] and the fitting error provides the error 
bar.

\begin{figure}[] \centering
\includegraphics[width=0.45\textwidth]{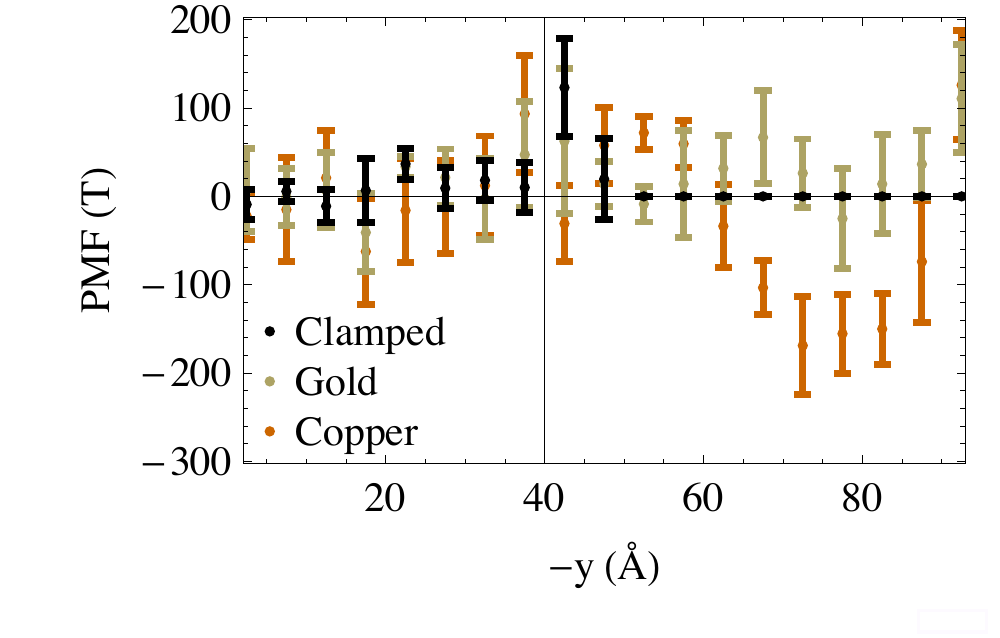}
\caption{(Color online) 
The PMF of graphene pressurized through equally sized circular apertures along a 
vertical section extending from the center of the aperture to the bottom of the 
simulation cell. The vertical line at $\sim 40$\,\AA\ marks the radius of the 
apertures. See the text in Appendix \ref{ap:sectional-PMF-circular} for 
more details.
}
\label{fig:circle_sectional}\end{figure}

\section{Sectional plot of PMFs for circular apertures}%
\label{ap:sectional-PMF-circular}

To better illustrate the magnitude of the PMF in the vicinity of apertures 
simulated under realistic substrate conditions, we present here a sectional 
view of the field for the representative cases of the circular apertures 
simulated in the artificially clamped, Au, and Cu scenarios explored in the 
main text. \Fref{fig:circle_sectional} shows the PMF of graphene pressurized 
through circular apertures of the same size sampled along a vertical section 
extending from the center of the aperture to the bottom of the simulation cell. 
The sections are taken from the corresponding data shown in 
Figs.~\ref{fig:allbubbles}(a), \ref{fig:allbubblesAu}(a), and 
\ref{fig:allbubblesCu}(a) by sampling the PMF along a vertical direction and 
performing averages within square bins of 25\,\AA$^2$. The averaging is done to 
account for local fluctuations in the PMF and the standard deviation in each 
bin is used to draw the error bars. The traces in \Fref{fig:circle_sectional} 
are analogous to the ones in Figs.~\ref{fig:circle_angular} or 
\ref{fig:PBD}(a), with the exception that there is no angular averaging here 
because the substrate interaction breaks the rotational symmetry [cf., for 
example, the region outside the aperture in \Fref{fig:allbubblesCu}]; 
consequently the error bars are higher here than in the artificially clamped 
cases.

\section{Comparison of PMFs from displacement and full TB approaches}%
\label{ap:pmf-disp}

As described in the main text, the \emph{displacement approach} to obtain the 
pseudomagnetic fields throughout the graphene sheet consists in 
directly employing \Eqref{eq:Adef-strain}, where the components of the strain 
tensor are extracted numerically from the MD-relaxed atomic positions. Apart 
from contributions beyond linear order in strain, this should be equivalent to 
computing the vector potential $\mbf{A}(\mbf{R})$ directly from the definitions 
\eqref{eq:Adef-full} and \eqref{eq:Hopping}, but neglecting the bending effects 
in the hopping. This amounts to considering $-t_{ij} = V_{pp\pi}(d)$.

\begin{figure*} \centering
\includegraphics[width=0.85\textwidth]{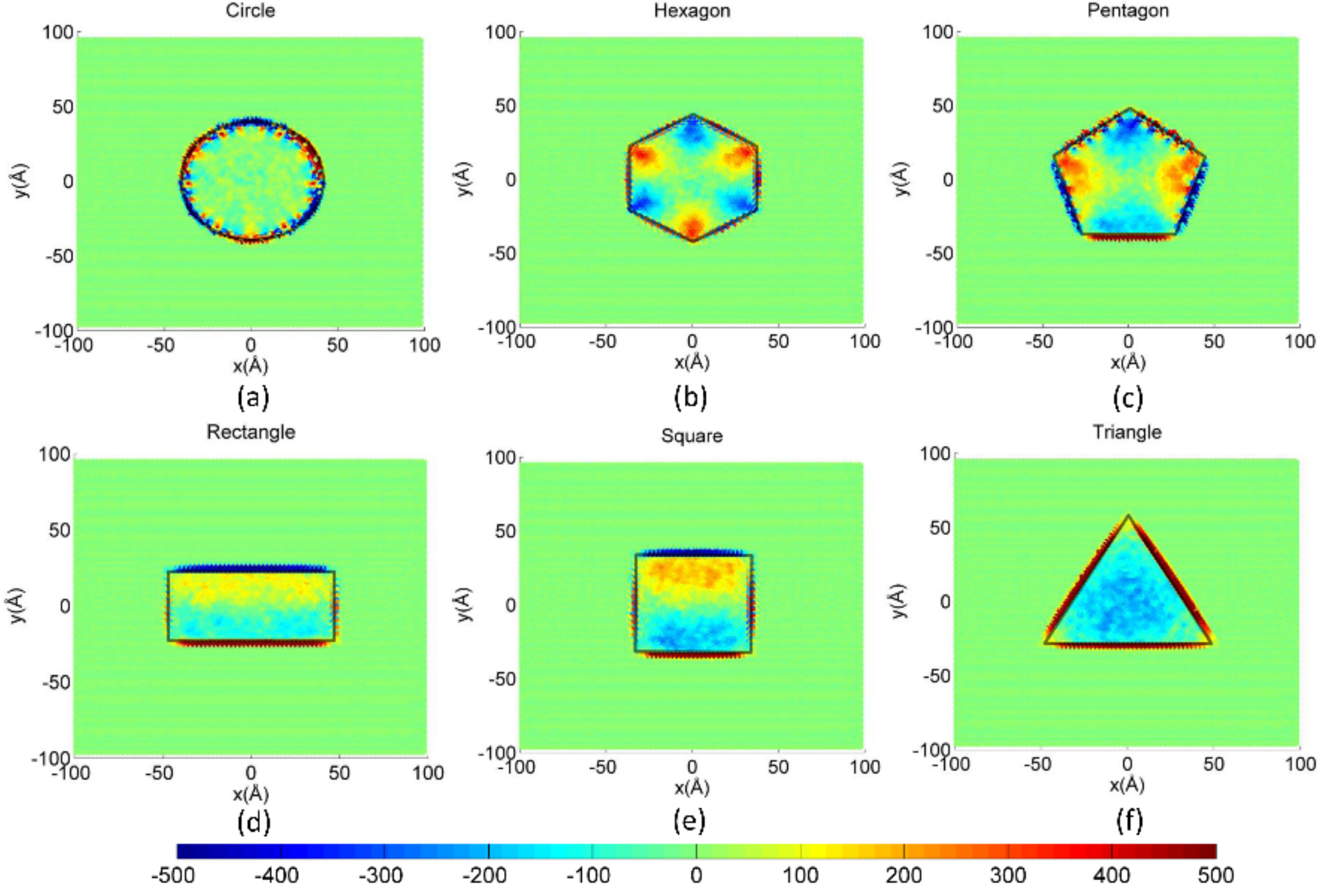}
\caption{(Color online) PMF distribution for the same systems analyzed in
\Fref{fig:allbubbles}, but here the field is computed by the displacement 
approach discussed in the main text. The PMF scale is in units of Tesla. The 
edge of the substrate apertures used in the MD simulations is outlined (gray 
line) for reference.
}
\label{fig:allbubbles-disp}\end{figure*}

For completeness, and to show that the two approaches lead to the same results 
in practice, we present in \Fref{fig:allbubbles-disp} the PMF distribution 
computed by the displacement approach for the same systems analyzed in 
\Fref{fig:allbubbles}. The agreement is very satisfactory and shows that the 
displacement and tight-binding methods are equivalent if curvature can be 
neglected.

\section{Comparison of Displacement and Stress Approaches}%
\label{ap:disp-vs-stress}

The final (inflated bubble) configuration gives us the basic ingredients needed 
to calculate the strain, \ie, the deformed atomic positions.  Here we 
present further details on the displacement and stress approaches we 
investigated for calculating the strain. In the end, the stress approach 
revealed itself inadequate to accurately capture the local strain in the 
graphene lattice.

\subsection{Displacement Approach}

We begin with the continuum definition for strain~\cite{ZimmermanIJSS2009}, 
which is written as
\begin{equation} \label{eq:straindisp-ap}
\epsilon_{ij}=\frac{1}{2}\left(\frac{\partial {u_i}}{\partial {X_j}} + 
\frac{\partial {u_j}}{\partial 
{X_i}}\right)+\frac{1}{2}\frac{\partial{u_k}}{\partial{X_i}}\frac{\partial{u_k}}
{\partial{X_j}},
\end{equation}
where $\epsilon_{ij}$ are the components of the strain, $u$ is the 
displacement, 
and $X$ denotes the position of a point in the reference configuration. To 
compute the displacement field that is needed to evaluate the strain in 
Eq.~\ref{eq:straindisp-ap}, we first exploit the geometry of the graphene 
lattice by 
meshing it using tetrahedral finite elements~\cite{hughes1987}, where each 
finite element is comprised of four atoms.  To remove spurious rigid body 
translation and rotation modes, we choose the deformed position of the atom of 
interest (atom 0) to be the reference position, \ie, 
$\mbf{r}_{0}=\mbf{R}_{0}$. 
 
By subtracting the original position of each neighboring atom from its deformed 
position, we obtain the displacement vectors of the three nearest neighbors: 
$\mbf{u_{01}}=(u_{01x},u_{01y},u_{01z})$, 
$\mbf{u_{02}}=(u_{02x},u_{02y},u_{02z})$, 
$\mbf{u_{03}}=(u_{03x},u_{03y},u_{03z})$. 

We use the linear interpolation property of the four-node tetrahedral element 
to 
denote the displacement field $\mbf{U}(x,y,z) = (U_{x},U_{y},U_{z})$ inside the 
tetrahedral element as: $U_x=a_1x+a_2y+a_3z$, $U_y=a_4x+a_5y+a_6z$, 
$U_z=a_7x+a_8y+a_9z$, where $a_1$ to $a_9$ are unknown constants for each 
tetrahedral element.  Inserting the positions ($\mbf{r_1}=(x_1,y_1,z_1)$, 
$\mbf{r_2}=(x_2,y_2,z_2)$, $\mbf{r_3}=(x_3,y_3,z_3)$) and the corresponding 
displacements ($\mbf{u_{01}}$,$\mbf{u_{02}}$,$\mbf{u_{03}}$) of the three 
neighboring atoms, we can solve $a_1$ to $a_9$ in terms of $\mbf{r_1}$, 
$\mbf{r_2}$, $\mbf{r_3}$ and $\mbf{u_{01}}$, $\mbf{u_{02}}$ and $\mbf{u_{03}}$, 
thus obtaining all coefficients of $\mbf{U}(x,y,z)$. If we rearrange 
$\mbf{U}(x,y,z)$ to express it in terms of $\mbf{u_{01}}$,$\mbf{u_{02}}$ and 
$\mbf{u_{03}}$, we obtain the following equation:
\begin{equation} \label{eq:dispwhole}
\left[
\begin{matrix}
  U_x \\
  U_y\\
  U_z\\
\end{matrix}
\right]
\!\!=\!\!
\left[
\begin{matrix}
  N_1 & 0 & 0 & N_2 & 0 & 0 & N_3 & 0 & 0\\
  0 & N_1 & 0 & 0 & N_2 & 0 & 0 & N_3 & 0\\
  0 & 0 & N_1 & 0 & 0 & N_2 & 0 & 0 & N_3\\
\end{matrix}
\right]
\!\!
\left[
\begin{matrix} 
  u_{01x}\\
  u_{01y}\\
  u_{01z}\\
  u_{02x}\\
  u_{02y}\\
  u_{02z}\\
  u_{03x}\\
  u_{03y}\\
  u_{03z}\\
\end{matrix}
\right]\!\!,
\end{equation}
where $N_i=N_i(x,y,z), i = 1, 2, 3$ are the finite element shape functions.  
For simplicity, we can rewrite Eq.~\eqref{eq:dispwhole} as:
\begin{equation} \label{eq:dispsimp}
\mbf{U}=\mbf{N}\bm{\cdot}\mbf{u_N},
\end{equation}
where $\mbf{u_N}=[\mbf{u_{01}}, \mbf{u_{02}}, \mbf{u_{03}}]^T$ is the 
displacement field of the three neighbor atoms.

After we obtain the displacement field $\mbf{U}$, the strain can be derived by 
differentiating Eq.~\eqref{eq:dispsimp} following the continuum strain as 
defined in Eq.~\eqref{eq:straindisp-ap} to give
\begin{equation} \label{eq:strainsimp}
\mbf{\epsilon}=\mbf{T}\bm{\cdot}\mbf{u_N},
\end{equation}
where $\mbf{T}=\frac{\partial \mbf{N}}{\partial \mbf{x}}$ is constant inside 
each tetrahedral element. Once the strains for each atom are determined, the 
vector gauge field $\mbf{A}$ is straightforward to compute.  However, to get 
the 
pseudomagnetic field, $B=\partial_x A_y - \partial_y A_x$, another derivative 
is needed, calculated in a 
similar fashion as the strain is calculated from the displacement field.  Thus, 
the displacement approach involves two numerical derivatives, but no 
approximation is made about material properties. 

\begin{figure*}
\begin{minipage}[b]{.48\textwidth}
  \centering
  \includegraphics[scale=0.35]{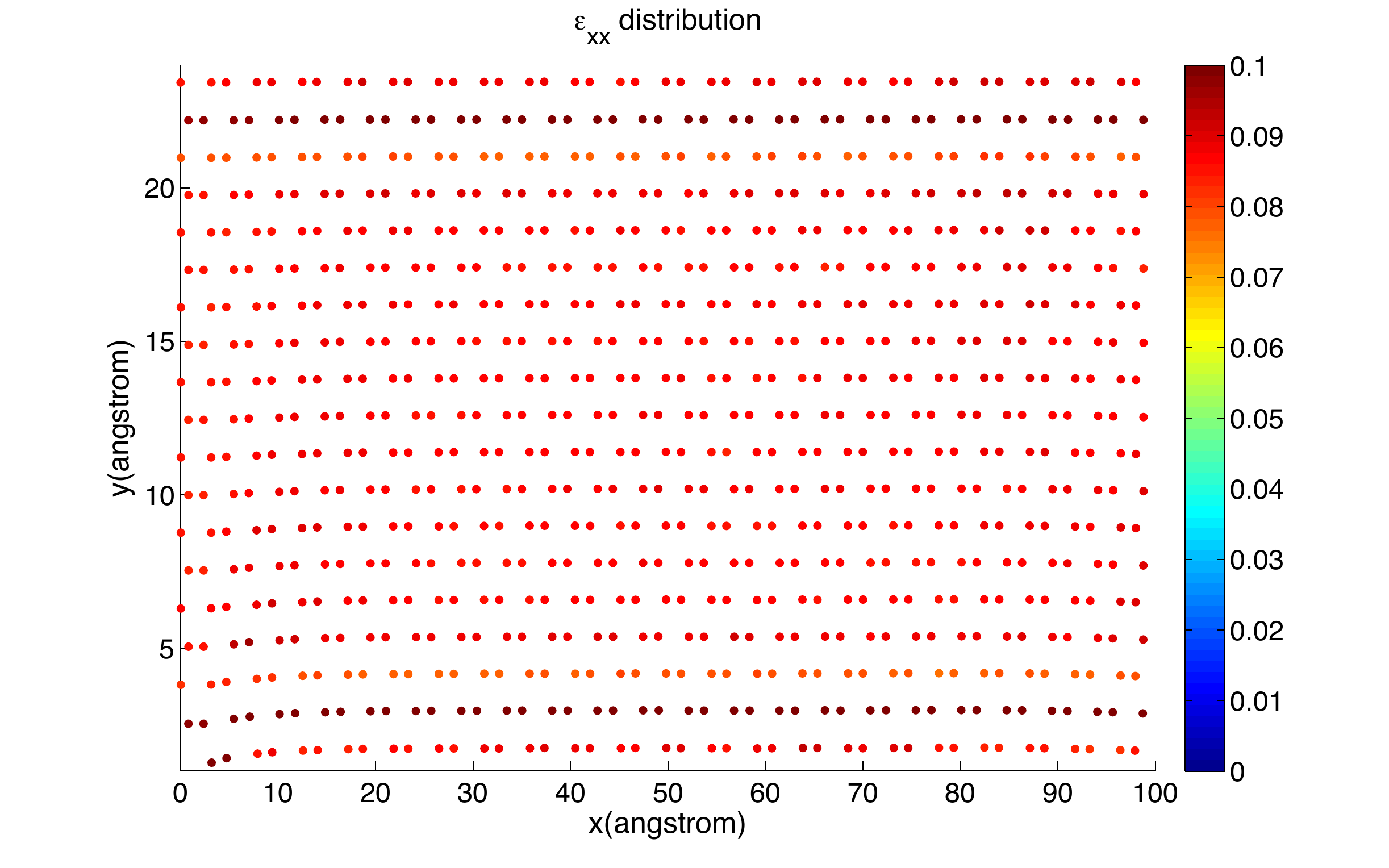}
  \caption{(Color online)  $\epsilon_{xx}$ distribution by displacement 
approach 
for uniaxial stretching case with 10\,\% strain. \label{S1}}
\end{minipage}
\hfill
\begin{minipage}[b]{.48\textwidth}
  \centering
  \includegraphics[scale=0.35]{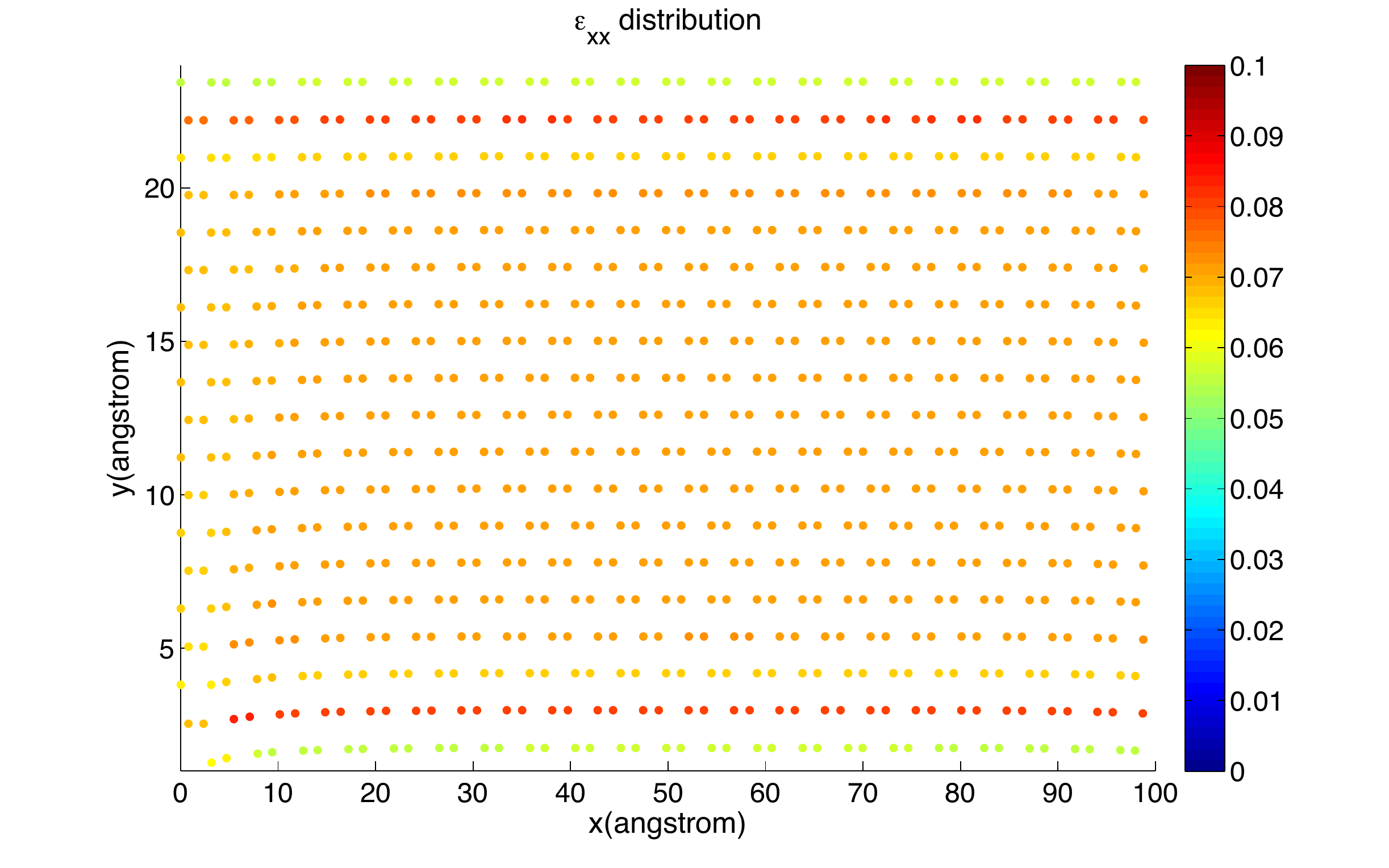}
  \caption{(Color online)  $\epsilon_{xx}$ distribution by stress approach for 
uniaxial stretching case with 10\,\% strain.
  \label{S2}}
\end{minipage}
\par\bigskip\par
\begin{minipage}[b]{.48\textwidth}
  \centering
  \includegraphics[scale=0.33]{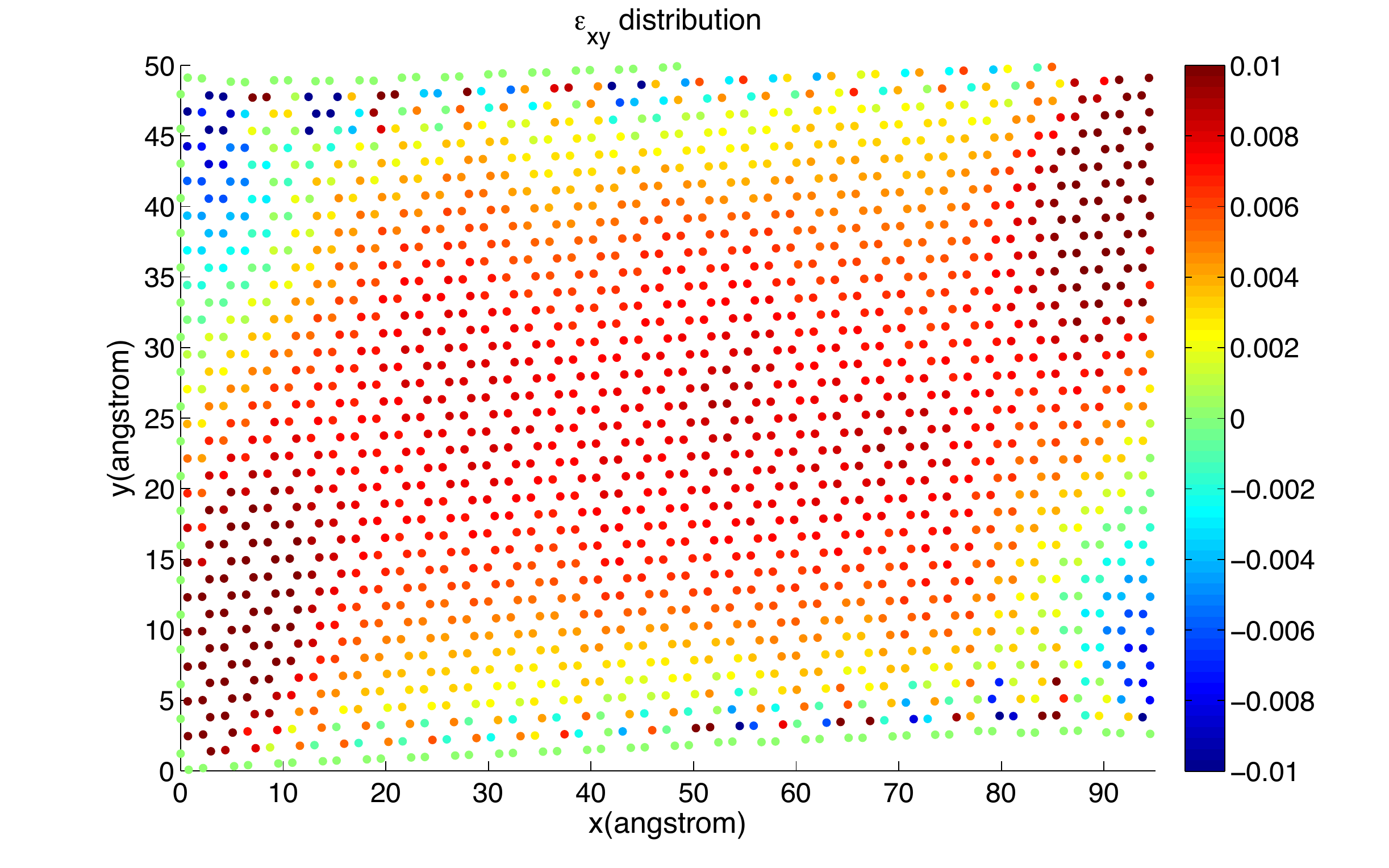}
  \caption{(Color online)  $\epsilon_{xy}$ distribution by displacement 
approach 
for simple shear case with 1\,\% strain.
  \label{S3}}
\end{minipage}
\hfill
  \begin{minipage}[b]{.48\textwidth}
  \centering
  \includegraphics[scale=0.33]{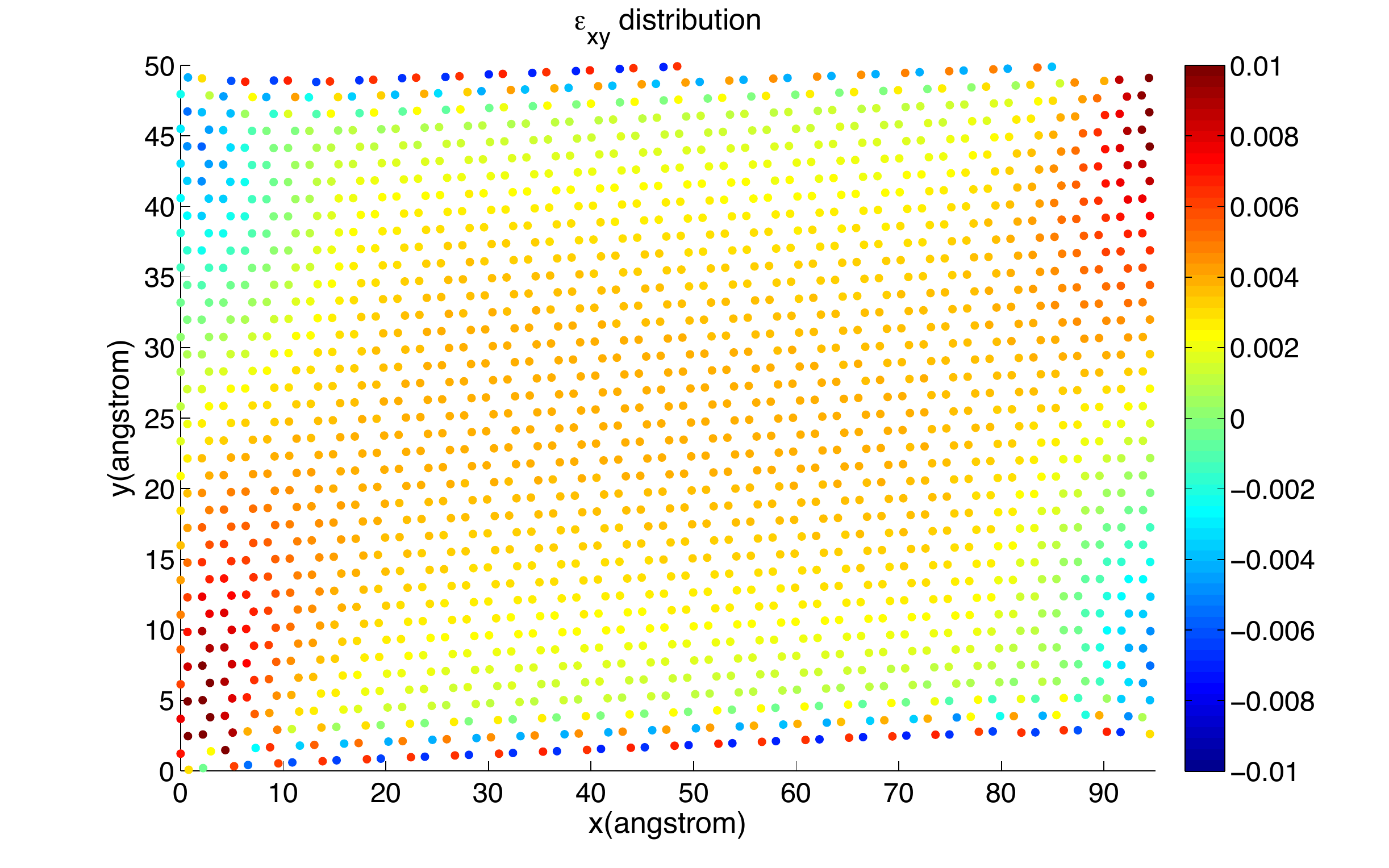}
  \caption{(Color online)  $\epsilon_{xy}$ distribution by stress approach for 
simple shear case with 1\,\% strain.
  \label{S4}}
\end{minipage}
\end{figure*}

\subsection{Stress Approach}

In MD simulations, the atomic virial stress can be extracted on a
per-atom basis.  In the present work, the virial stress as calculated
from LAMMPS~\cite{ThompsonJCP2009} was obtained for the final
(inflated) graphene bubble configuration.  These stresses were then
related to the strain via a linear constitutive relationship, as was
done recently by~\citet{KlimovScience2012}. In the current work, we
utilized a plane stress model for graphene, where the in-plane
strains are written as
$\epsilon_{xx}=\frac{1}{E}(\sigma_{xx}-\mu\sigma_{yy})$,
$\epsilon_{yy}=\frac{1}{E}(\sigma_{yy}-\mu\sigma_{xx})$,
$\epsilon_{xy}=\frac{\sigma_{xy}}{G}$.  The material properties of
graphene are chosen as $E$ = 1\,TPa~\cite{huangPRB2006}, $G$ =
0.47\,TPa~\cite{MinAPL2011} and $\mu$ = 0.165~\cite{BlaksleeJAP1970},
where $E$ is the Young's modulus, $G$ the shear modulus, and $\mu$
Poisson's ratio. It is important to note that, because a linear
stress-strain relationship is assumed, the resulting strain is
generally {\it underestimated}, particularly at large deformations due
to the well-known nonlinear stress-strain response of
graphene~\cite{leeSCIENCE2008}.

Both potential and kinetic parts were taken into account for virial stress 
calculation.  We note that the virial stress calculated in LAMMPS is in units 
of 
``Pressure$\,\cdot\,$Volume", and thus we used the standard value of 3.42\,\AA \
as the effective thickness of single layer graphene~\cite{huangPRB2006} to 
calculate the stress.  A plane stress constitutive model was utilized to 
calculate the strain via
\begin{equation} \label{eq:planestress}
\left[
\begin{matrix}
\epsilon_{xx}\\
\epsilon_{yy}\\
\epsilon_{xy}\\
\end{matrix}
\right]
=
\left[
\begin{matrix}
\frac{1}{E} & -\frac{\mu}{E} & 0\\
-\frac{\mu}{E} & \frac{1}{E} & 0\\
0 & 0 & \frac{1}{2G}\\
\end{matrix}
\right]
\cdot
\left[
\begin{matrix}
\sigma_{xx}\\
\sigma_{yy}\\
\sigma_{xy}\\
\end{matrix}
\right],
\end{equation}
where the constitutive parameters are given in the main text of the manuscript. 
 
After the strain is obtained, the same method as in the displacement approach 
was used to calculate the vector gauge field $\mbf{A}$ and the pseudomagnetic 
field $B$. The stress approach avoids one numerical differentiation but a 
constitutive approximation is involved, \ie,  that the stress-strain 
response for graphene is always linear.

\subsection{Benchmark Examples}

We compare the displacement and stress approaches via two simple benchmark 
examples, those of uniaxial stretching and simple shear.  For the uniaxial 
stretching case, $\epsilon_{xx}\approx10\,\%$ strain was applied along the 
x-direction.  The loading was done by applying a ramp displacement that went 
from zero in the middle of simulation box to a maximum value at the +x and -x 
edges of the graphene monolayers. 

For the simple shear case, $\epsilon_{xy}\approx1\,\%$ shear strain was applied 
by fixing the -x edge and displacing the +x edge in the y-direction.  Both the 
uniaxial stretching and simple shear simulations were performed via classical 
MD 
simulations using the open source LAMMPS~\cite{plimptonLAMMPS} code with the 
AIREBO potential~\cite{stuartJCP2000}.  The result for the uniaxial stretching 
is shown in Figs.~\ref{S1}\ and Fig.~\ref{S2}, while the simple shear is shown 
in Figs.~\ref{S3}\ and Fig.~\ref{S4}.  The superior performance of the 
displacement approach is seen in both cases.  Specifically, because a linear
stress-strain relationship is assumed in the stress approach as shown in 
Eq.~\eqref{eq:planestress}, the resulting strain is generally {\it 
underestimated}, particularly at large deformations due to the well-known 
nonlinear stress-strain response of graphene~\cite{leeSCIENCE2008}. 

Once the strain distribution is determined from the MD simulations the
PMF, $B$, can be directly evaluated from the definitions above.
However, if the strain tensor is calculated within the deformation
approach, a second numerical derivative is needed to get $B$, which
is likely to introduce a certain degree of error. Nevertheless we
found the errors to be of acceptable magnitude.

Compared with the displacement approach, the stress approach avoids
one numerical differentiation, but a constitutive approximation is
involved.  To compare the accuracy of the displacement and stress
approaches, we calculated the PMF distribution in a circular bubble
(for which an analytic solution is available and detailed analysis was
recently performed~\cite{KimPRB2011}) by obtaining the strain via
three different methods, as illustrated in Fig.~\ref{fig:circle}: an
analytic continuum mechanics model, \ie,  the Hencky
solution~\cite{FichterNASA1997} (b), the MD-based displacement approach (c),
and the MD-based stress approach (d).  In the MD simulations we used
100 snapshots over 5\,ps during thermal equilibrium to determine the
average final position and stress for the inflated bubbles.  For all
three models, the radius of the circular hole was 3\,nm, while the
final deflection was about 1\,nm.

As Fig.~\ref{fig:circle} demonstrates, the PMFs generated from
the MD-based displacement approach are in good agreement with those
that follow from Hencky's analytic solution, and also with previously
reported values for a circular bubble~\cite{KimPRB2011}. In contrast,
the stress approach fails to yield reasonable results for this loading
situation, even at the qualitative level.

\bibliography{biball}

\end{document}